\documentclass[aps,article,nofootinbib]{revtex4}

\usepackage{graphicx}
\usepackage{amsfonts}
\usepackage{amsmath}
\newcommand{\be}{\begin{eqnarray}}
\newcommand{\ee}{\end{eqnarray}}

\newcommand{\ox}{\overline{x}}

\begin{document}
\title{Characterization of Protein Folding by Dominant Reaction Pathways}

\author{Pietro Faccioli}
\email{faccioli@science.unitn.it}
\affiliation{Dipartimento di Fisica  Universit\'a degli Studi di Trento and I.N.F.N., Via Sommarive 14, Povo (Trento), I-38050 Italy.} 

\begin{abstract}

We assess the reliability of the recently developed approach denominated Dominant Reaction Pathways (DRP) by studying the folding of a 16-residue $\beta$-hairpin, within a coarse-grained Go-type model. We show that the DRP predictions are in quantitative agreement with the results of Molecular Dynamics simulations,  performed in the same model. On the other hand, in the DRP approach, the computational difficulties associated to the decoupling of time scales are rigorously bypassed. The  analysis of the important transition pathways supports a picture of the  $\beta$-hairpin folding in which the reaction is initiated by the collapse of the hydrophobic cluster.
\end{abstract}

\maketitle
\section{Introduction}
\label{Intro}

In this work we address the problem of identifying  the statistically important pathways in thermally activated conformational transitions of macro-molecules.
If such  reactions involve  overcoming free-energy barriers,  straightforward molecular dynamics (MD) simulations are very inefficient and often impracticable, due to the decoupling of the microscopic time scale associated to local conformational changes and the macroscopic time scale related to the inverse reaction rate. For example, in the case of protein folding, MD simulations can only access times of few $ns$, whereas the entire folding processes generally take milliseconds or longer. As a consequence of the existence of free-energy barriers between folded and unfolded states, all the computational time in  MD simulations  is invested for describing the motion of the system, when it is exploring the portion of configuration space associated to the stable and meta-stable states. 

In a series of recent works~\cite{DFP1,DFP2,DRP} , we have developed a theoretical/computational  framework, which we have called Dominant Reaction Pathways  (DRP), to rigorously identify the statistically significant pathways in thermally activated reactions of systems with a large number of degrees of freedom. Unlike other approaches  in the literature, the DRP method does not involve any uncontrolled {\it ad hoc} approximation and does not require any {\it a priori} choice of reaction coordinates. 
  
Before applying the DRP method to make quantitative predictions for complex molecular transitions, we need to show that it  produces results consistent with those of  MD simulations, obtained in the same model.
A first validation analysis  of this of this type was performed in \cite{DFP2}, where we studied the conformational transitions of a very simple molecule  ---alanine dipeptide--- using the all-atom {\it GROMOS96} force field. We observed an excellent agreement with the results of MD simulations, performed in the same model. On the other hand, the computational gain of adopting the DRP method  was impressive:   the characterization of the relevant transitions in atomistic detail took just few minutes on a regular desktop, while about a week of calculations on the same machine was required, in order to extract the same amount of information, from the standard MD techniques.
These results suggest that the  DRP method can be an efficient tool to characterize isomerizations, allosteric transitions and in general reactions  connecting  well-defined stable molecular conformations . 

On the other hand, if either the initial or the final state has  a very large conformational entropy, the characterization of the reaction becomes much more complex. A prototype of this class of problems is again the study of protein folding. 
Proteins can fold in an uncountable infinity of ways, corresponding to the infinite set of  denatured configurations form which the reaction can be initiated.
In this case, in order to extract significant information from either MD simulations or  DRP calculations,  one necessarily needs to perform an average over  a representative set of transitions, which reach the same native state starting from different denatured configurations. 
Hence, the question arises whether one needs  to average  over an exponentially large number of folding trajectories 
 or if a finite --- and possibly relatively small--- number of independent trajectories are sufficient to characterize the folding.
Clearly, in the first scenario, any approach which can at most provide a handful of folding transitions  looses its practical utility.

In this work,  we address this issue  by studying the folding of a 16-residue chain  ----$\beta$-hairpin fragment of GB1---, using a Go-type reduced model. 
The dynamics of the Go model is sufficiently simple to allow many folding-unfolding trajectories  to be generated by MD simulations, at an affordable computational cost. 
On the other hand, this model displays some of the complexity which we expect to face, in more realistic and detailed all-atom simulations of protein folding.  
In particular,  the reaction takes place in a high-dimensional funneled energy landscape in which the native state has a  small conformational entropy, while the denatured conformations visit a large portion of configuration space. 

The direct comparison with the results of  simulations shows that the DRP approach remains accurate and computationally efficient also for protein folding, at least for the simple model used. 
In fact, we can accurately predict the configurations which are most probably visited during the unfolding/refolding reaction, by averaging over the dominant reaction pathways corresponding to  just few initial denatured conformations.

In the final part of this work,  we study the folding mechanism for the hairpin under consideration. We find that our analysis favors a picture in which the folding is initiated by the collapse of an hydrophobic cluster. This result agrees with that of MD simulations performed in the same model and with most of the existing theoretical analysis, both in the context of coarse-grained and all atom models. Interestingly, florescence experiments   suggest a different mechanism, in which the folding is initiated at the turn (see \cite{Eaton1, Gai} and references therein).

The paper is organized as follows. In section \ref{model}, we introduce the Go-type model used in the MD and DRP calculations. In section \ref{inDRP}, we briefly review the DRP approach ---for a detailed and self-contained presentation, we refer the reader to  \cite{DRP}---, while some general and qualitative aspects of the stochastic dynamics in such a model are discussed in \ref{dDFP}. In section \ref{calculations} we present the algorithm used to characterize the ensemble of statistically significant folding transitions, including the effects of thermal fluctuations. 
The DRP results for the folding reaction of the $\beta$-hairpin are presented in section \ref{results}, where they are compared with the predictions of MD simulations. 
Section \ref{mechanism} is devoted to a discussion of the qualitative folding mechanism, which emerges from our DRP and MD calculations. The main results and conclusions are summarized in section \ref{conclusions}.

\section{Go-Type Model for the Folding of a $\beta$-Hairpin}
\label{model}
 \begin{figure}[t!]
	\includegraphics[clip=,width=10 cm]{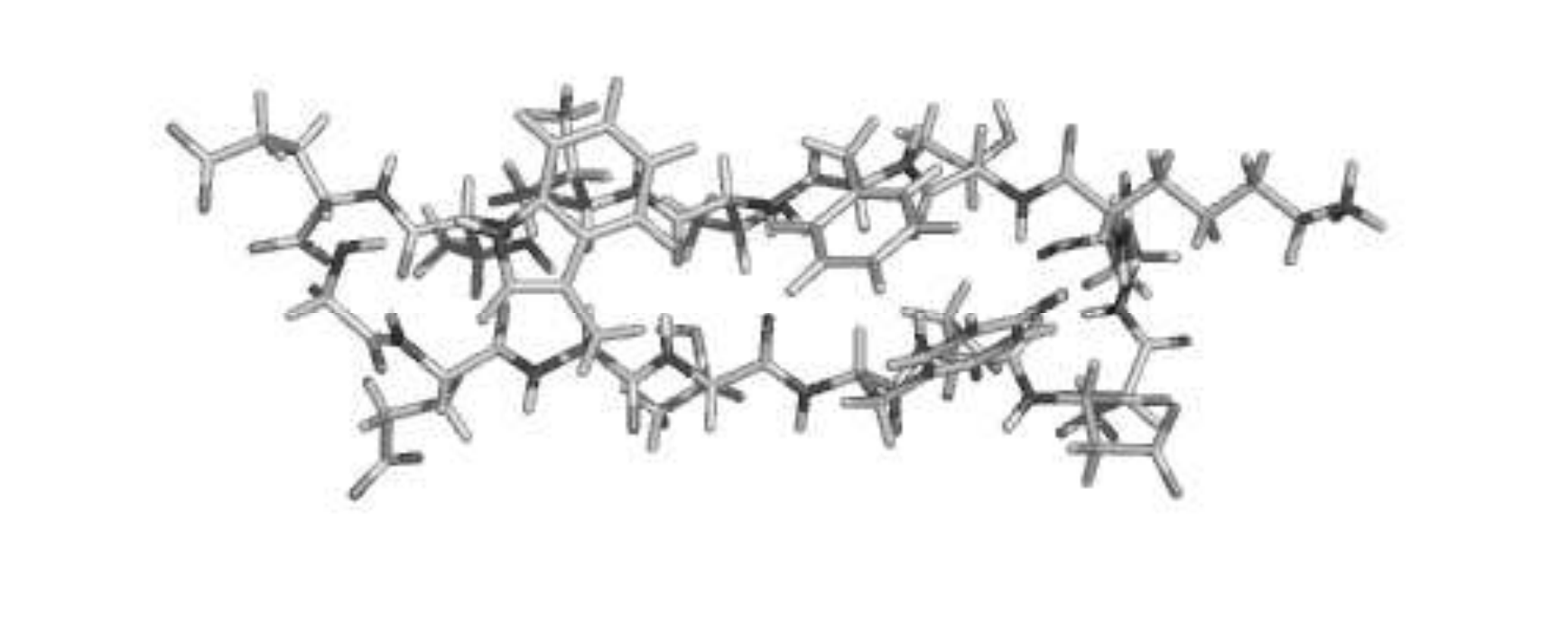}		
	\caption{16-residue C-terminus of protein G-B1 (PDB code 2gb1).}
	\label{GB1}
\end{figure}

In this work, we consider the folding reaction of the 16-residue C-terminus of protein G-B1 (PDB code 2gb1), with sequence
\be
\textrm{GLY-GLU-TRP-THR-TYR-ASP-ASP-ALA-THR-LYS-THR-PHE-THR-VAL-THR-GLU}
\nonumber
\ee
In solution, this fragment is known to assume the stable $\beta$-hairpin conformation shown in Fig. \ref{GB1}.

Fluorescence experiments performed by Eaton and
coworkers \cite{Eaton1} have shown that this hairpin exhibits two-state
kinetics between the folded and unfolded state, with a relaxation
time of 6 $\mu s$.
In the last decade, this peptide has become a standard test system, to investigate the thermodynamics and the kinetics of $\beta$-sheet formation (see e.g. \cite{GB1ref} and references therein). 

In order to be able to compare the results of MD and DRP calculations, we must adopt a model which is sufficiently simple to be investigated with either methods, at an affordable computational cost.
Hence, we consider a Go-type coarse-grained representation~\cite{Go}, defined by the interaction
\be
U = \sum_{i<j=1}^{N_p} u({\bf x}_i,{\bf x}_j)= \sum_{i<j=1}^{N_p}~\frac{1}{2} K_b \left(|{\bf x}_i-{\bf x}_{j}|-a\right)^2\delta_{j,i+1} 
+ 4~\epsilon \left[ \left(\frac{\sigma}{r_{ij}}\right)^{12} - C_{i j} \left(\frac{\sigma}{r_{ij}}\right)^6\right](1- \delta_{j,i+1}),
\label{Ugo}
\ee  
where $N_p=16$ is the number of amino-acids, $a=0.38\,nm$ is the equilibrium distance between consecutive $\alpha$-carbons on the chain, $\sigma= 0.25 nm$ is the residue's size\footnote{Since we were interested in the validation of the DRP method, and not on the phenomenological implications, in this work we chose to adopt a rather small hard-core radius, in order to make numerical simulations faster. For example, the acceptance rate of Monte Carlo simulations varies significantly with $\sigma$.}  $\epsilon= 5 kJ/mol$ is the maximum energy gain per contact and $K_b=2\times 10^3 kJ nm^{-2}/mol$ is the elastic constant of the harmonic bond.
The matrix $C_{i j}$ selects the native contacts: it is set equal to 1 if the native distance between residues $i$ and $j$ is less than $0.65~nm$, and is set to 0 otherwise.

In this  study, we chose to keep the interaction as simple as possible and do not include  Coulomb, angular or torsional terms in (\ref{Ugo}).  The energetic bias toward the native state is effective only when the monomers come  close, since the non-bonded interaction in (\ref{Ugo}) becomes negligible for distances greater than approximatively one $nm$.

\begin{figure}
		\includegraphics[clip=,angle=-90,width=5 cm]{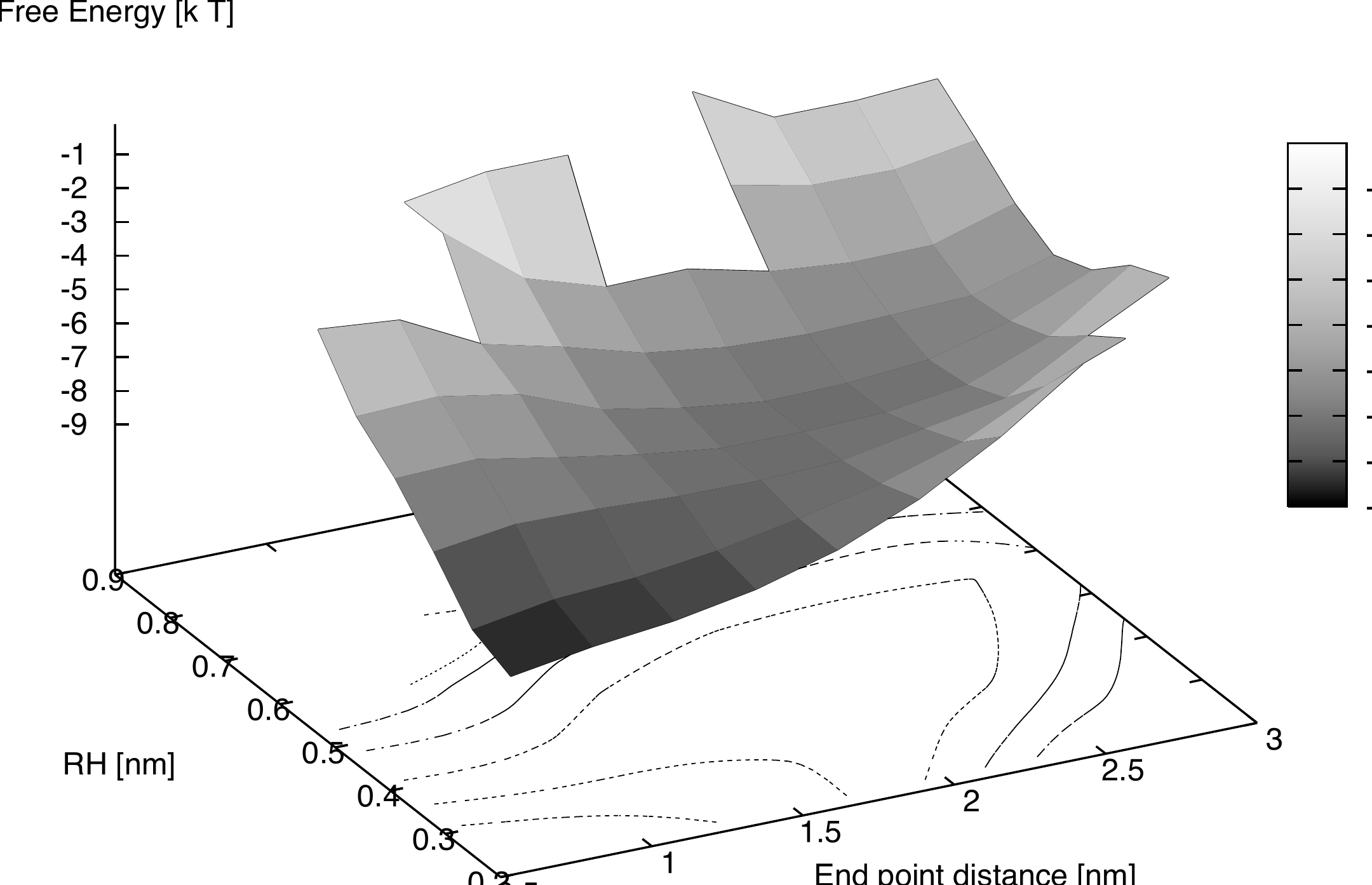}\vspace{0.3 cm}\hspace{0.5cm}
		\includegraphics[clip=,angle=-90,width=5 cm]{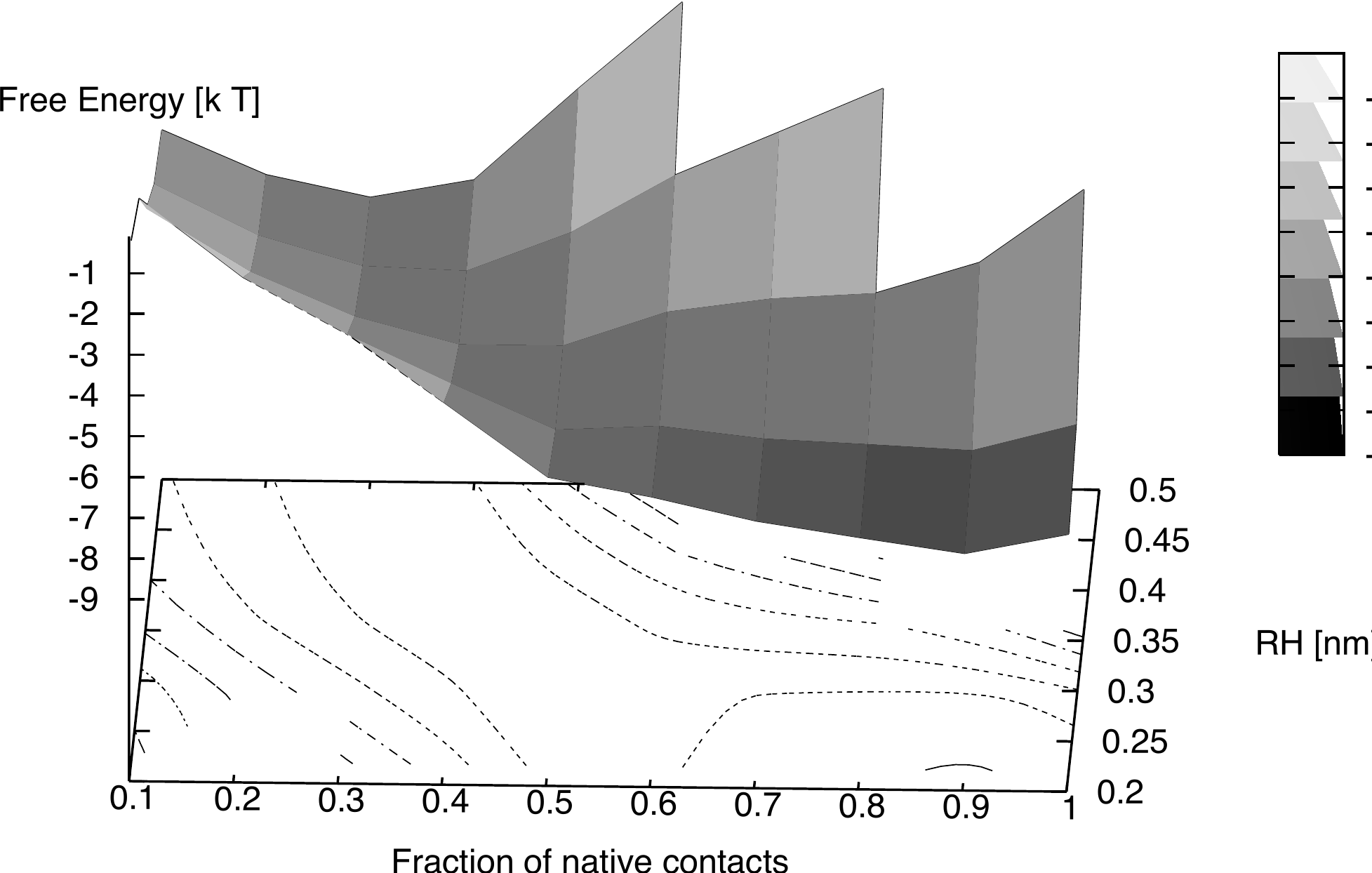}\vspace{0.3 cm}\hspace{0.5cm}
		\includegraphics[clip=,angle=-90,width=5 cm]{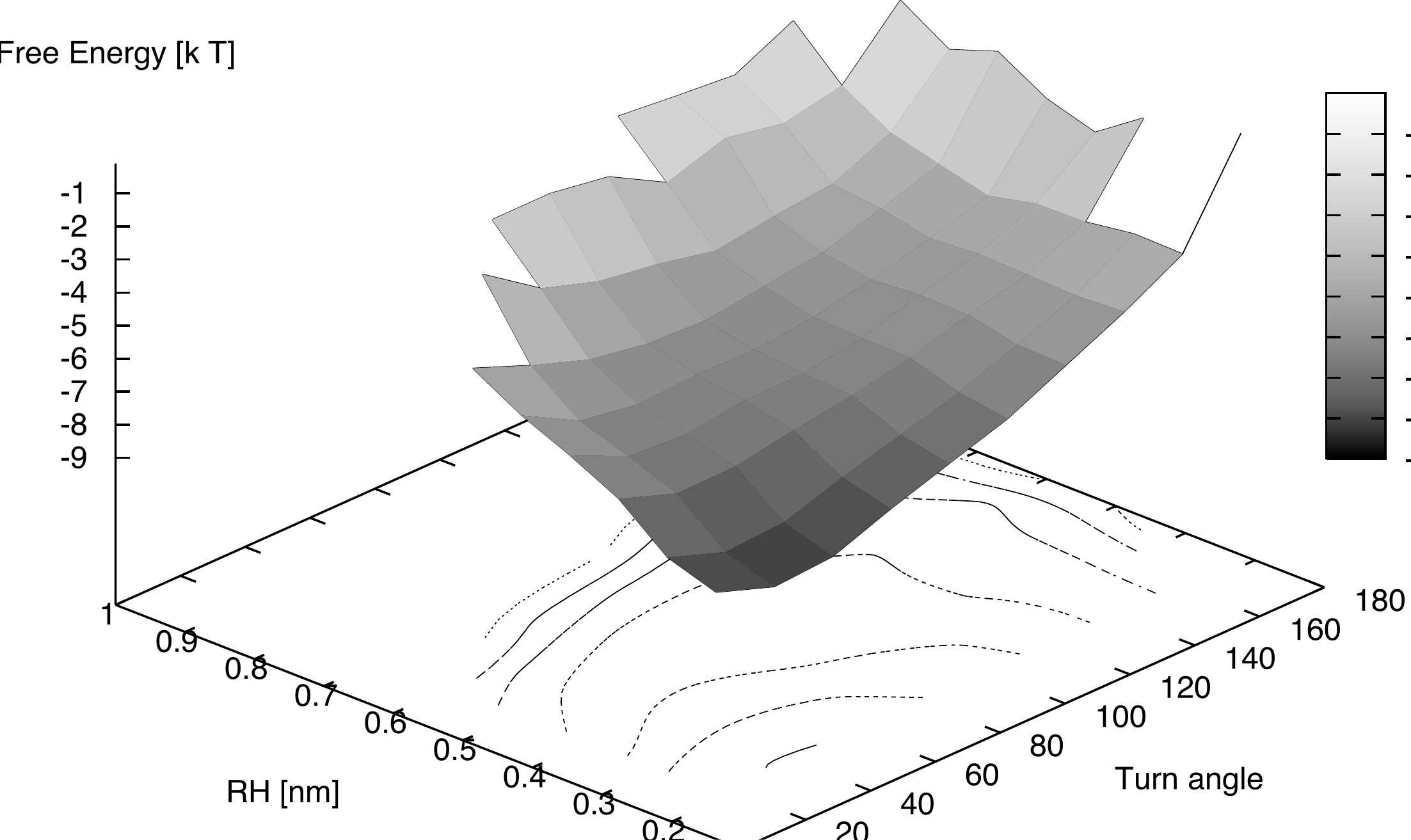}
	\caption{Free energy landscape in three surfaces identified by the parameters $d_{1-16}$,  $R_H$, and $N_c$. }
\label{G}
\end{figure}

The dynamics of the hairpin in a thermalized heat-bath can be described by 16 coupled  Langevin Eq.s.
\be
{\bf x}_1 &=& -\frac{D}{k_B T} {\bf \nabla}_1 U({\bf x}_1,...,{\bf x}_{N_p}) + {\bf \eta}_1(t),\nonumber\\
 &...&\nonumber\\
{\bf x}_{N_p} &=& -\frac{D}{k_B T} {\bf \nabla}_{N_p} U({\bf x}_1,...,{\bf x}_{N_p}) + {\bf \eta}_{N_p}(t).
\label{L}
\ee
where $\eta_i(t)$ are usual Gaussian noise functions, obeying the fluctuation-dissipation relationship
\be
\langle {\bf \eta}_{i}(t){\bf \eta}_{j}(0)\rangle = 6 D \, \delta(t) \delta_{i j},
\ee 
and $D= 1.2 \times 10^{-3} nm^2 ps^{-1}$ is the diffusion coefficient of a typical amino-acid in water, at room temperature. Notice that in the original Langevin Eq. there is an acceleration term,
$m \ddot{x}$. However,
 as it was shown in \cite{orland}, for proteins, the effect of including such a  term  becomes  negligible for time scales larger than the fraction of the $ps$.

\begin{figure}
		\includegraphics[angle=-90,width=8 cm]{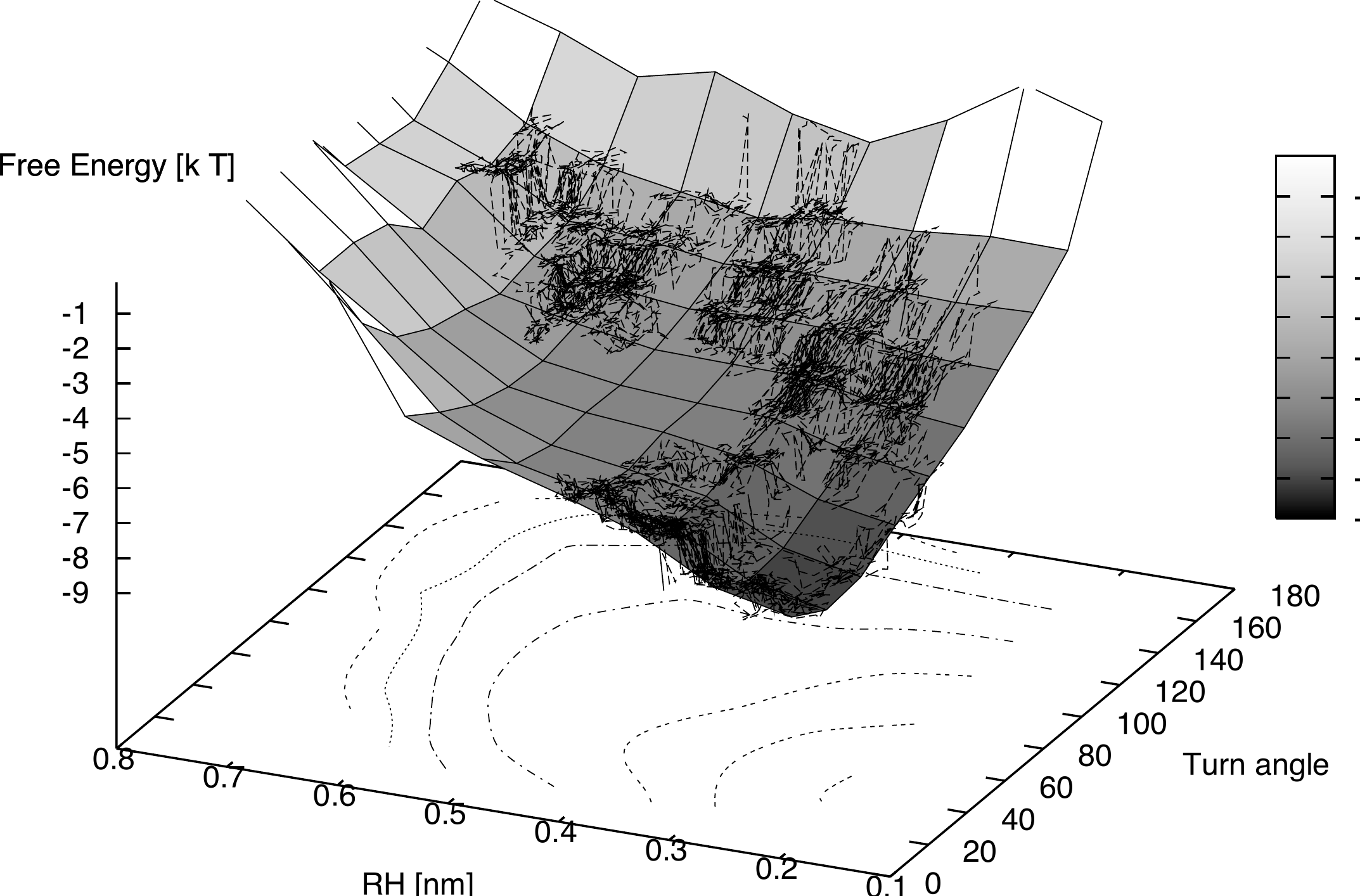}
	\caption{ An example of folding transition observed during a MD simulation, plotted in the plane selected by the hard-core radius  $R_H$ and the turn angle $\phi$}
\label{GMD}
\end{figure}

\begin{figure}
		\includegraphics[clip=,width=8 cm]{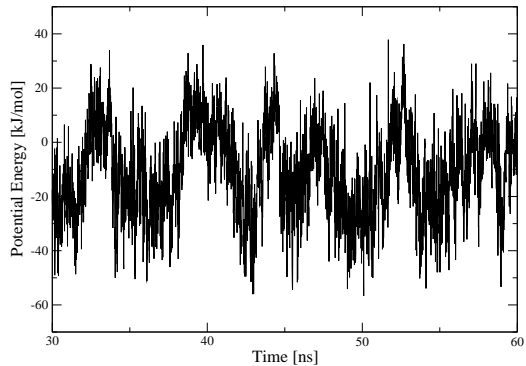}
	\caption{Time evolution of the potential energy during a fraction of a MD simulation.}
	\label{MD2state}
\end{figure}

Let us now discuss the key features of the dynamics of such a model, which can be inferred from performing  numerical MD simulations \footnote{In this work, we shall improperly call MD simulation the numerical integration of the Langevin Eq. (\ref{L}) in the so-called "Ito Calculus".} .
To this end, let us focus on the following set of order parameters:
\begin{itemize}
\item The distance $d_{1-16}$ between the $C$ and $N$ terminus of the chain
\item The angle $\phi$ at the turn of the hairpin, i.e. between the {\it Asp}, {\it Ala} and {\it Lys} residues, 
\be
\phi= \arccos \left[\frac{d_{6-8}^2+d_{8-10}^2-d_{6-10}^2}{2 d_{6-8}^2 d_{8-10}^2 } \right],
\ee
where $d_{i-j}$ is the distance between the $i$-th and $j$-th residues in the sequence.
\item The fraction of native contacts 
\item The radius of gyration $R_H$ of the group defined by the hydrophobic residues {\it Phe}, {\it Tyr}, and {\it Trp}.
\end{itemize} 
We stress the fact that the choice of this specific set of  parameters in only functional to the interpretation of our results. Neither MD simulations nor the  DRP method require any {\it a priori} choice of reaction coordinates.
Fig. \ref{G} displays the free energy landscape in three sufaces selected by the above order parameters, calculated  from  $\simeq 10^8$ steps of MD, with an integration time-step of $0.01 ps$.
An example of a typical folding transition observed during a MD simulation is displayed in Fig.\ref{GMD}, in the plane selected by the hard-core radius  $R_H$ and the turn angle $\phi$.
In Fig.\ref{MD2state} we show the evolution of the potential energy of the chain, during a  $30 ns$ long fraction of an MD simulation, at room temperature. 

These results show that  this Go-type model displays a unique, thermodynamically stable fold. The unfolded configurations do not form a thermodynamically stable state, and the folding reaction proceeds downhill.  A closer look on the three free energy profiles in Fig.\ref{G}   reveals that in each profile there is a region in which the free energy is nearly flat.
The corresponding region of configuration space is that in which the denatured chain spends  the longest time. 
Hence, the most probable unfolding reaction pathways are those which leave the native state, travel along the valley of the free energy profile and reach the regions where the free energy is almost flat.
In  section \ref{results} we shall see that the DRP method is able to predict and accurately locate such most probable trajectories, without any prior knowledge of the free energy profile.

\section{The DRP Approach}
\label{inDRP}

The starting point of the DRP approach is the path integral representation of the Fokker-Planck conditional probability to perform a transition from an initial  configuration $x_i$ to a final configuration $x_f$ in a time $t$, i.e. 
\be
P(x_f,t|x_i,0)= e^{-\frac{(U(x_f)-U(x_i))}{2 k_B T}}~\int_{x_i}^{x_f} 
\mathcal{D}x(\tau)~e^{- \int_{0}^{t} d\,\tau~ \left(\frac{\dot{x}^2(\tau)}{4 D}+ V_{eff}[x(\tau)]\right)}.
\label{path2}
\ee
$D$ is a constant diffusion coefficient, and  $V_{eff}(x)$ is an effective potential defined as
\be
V_{eff}(x)= \frac{D}{4(k_B T)^2} \left((\nabla U(x))^2 - 2 k_B T \nabla^2 U(x) \right),
\label{Veff}
\ee
$x=({\bf x}_1,...,{\bf x}_{N_p})$ is a vector specifying the position of all the $N_p$ constituents of the molecule (atoms or amino-acids). 
In the specific case of the study of the protein folding reaction,   the initial configuration $x_i$ may be taken  in the denatured state, while the final configuration $x_f\equiv x_N$ lies in native state.
 
The DRP approach is based on the  saddle-point approximation of  the path integral (\ref{path2}).  The idea is to functionally expand  around the minima of the effective action $S_{eff}=\int_{0}^{t} d\,\tau~ \left(\frac{\dot{x}^2(\tau)}{4 D}+ V_{eff}[x(\tau)]\right)$. The minimum action paths are the most probable trajectories ---or {\it dominant reaction pathways}--- which were first discussed in \cite{DFP1, DFP2} and \cite{DRP}.  
The key advantage of the DRP method is that the dominant reaction pathways can be very efficiently determined,  by exploiting the fact that they are solutions of a symplectic classical equations of motion, 
\be
\frac{1}{2D} \ddot x(\tau)= \nabla V_{eff}(x).
\label{DFPeom}
\ee
The solutions of  (\ref{DFPeom}) conserve an effective "energy" $E_{eff}$ defined as\footnote{Note that $E_{eff}$  has the dimension of a rate and therefore does not have the physical interpretation of a mechanical energy.}
\be
E_{eff}= \frac{1}{4D} \dot{x}^2 - V_{eff}(x).
\ee
As a consequence, the action $S_{eff}$ evaluated along a dominant reaction pathway  can be re-written as 
\be
S_{eff}(x_f,x_i, t) = - E(t) t +  S_{HJ}(x_f,x_i; E_{eff}(t)),
\ee
where $S_{HJ}(x_f, x_i; E_{eff})$ is the Hamilton-Jacobi (HJ) functional defined as
\be
S_{HJ}(x_f, x_i;E_{eff}(t)) = \frac{1}{\sqrt{D}}~\int_{x_i}^{x_f}d l \sqrt{E_{eff}(t) + V_{eff}[\ox_i(l)]},
\label{SHJ}
\ee
where $dl = \sqrt{d x^2}$ measures the infinitesimal displacement  of  all  the constituents in an elementary path step. 

Each choice of the effective energy selects the time taken by a single transition, from the initial configuration $x_i$ to the final configuration $x_f$, according to the well known relationship
\be
t= \int_{x_i}^{x_f} dl ~\frac{1}{\sqrt{4 D~(E_{eff}(t)+V_{eff}[\ox_i(l)])}}.
\label{time}
\ee
In a protein folding transition, the final state $\ox(t)$ should be taken as the native state of the protein $\ox_f$, that is the absolute minimum of $U(x)$, or possibly a minimum of $V_{eff}(x)$ close to it. In order for the protein to stay in this state as long as possible, the velocity of the system at that point should be $\dot \ox(t) = \dot \ox_f =0$, which implies
 \be
E_{eff}= \frac{1}{4D}\dot{\ox}^2(t)-V_{eff}(\ox(t))=-V_{eff}(\ox(t)).
\label{Efft}
\ee

The dominant reaction pathways are practically found by minimizing numerically a discretized version of Eq. (\ref{SHJ})
\be
S_{HJ}=\frac{1}{\sqrt{D}}~\sum_n^{N_s-1}\sqrt{\left(V_{eff}(n)-V_{eff}(1)\right)} \Delta l_{n,n+1}+\lambda P,
\label{DSHJ}
\ee 
where  $N_s$ is the total number of discretization steps and
\be
V_{eff}(n)&=&\frac{D}{4 (k_BT)^2} \sum_{i=1}^{N} \left[  \left(\sum_{j=1}^{N} {\bf \nabla}_j U({\bf x}_1(n),...,{\bf x}_N(n))\right)^2 -2 k_B T \sum_{j=1}^{N_p} \nabla^2_j U({\bf x}_1(n),...,{\bf x}_N(n))\right]\\
P&=&\sum_n^{N_s-1} (\Delta l_{n,n+1}-\langle \Delta l\rangle)^2,\\
(\Delta l)^2_{n,n+1}&=&\sum_{i=1}^{N_p}({\bf x}_i(n+1)-{\bf x}_i(n))^2.
\ee
$N_p$ is the number of particles (atoms  or residues) in the molecule, $\Delta\,l_{n,n+1}$ is the Euclidean measure of the $n-th$ elementary path-step and
$\lambda~P$ is a penalty function  which keeps all the length elements close to their average~\cite{Elber1} and becomes irrelevant in the continuum limit, $N\to\infty$.  For any finite number of path discretization steps,  this term enforces the condition that the relevant transitions are sampled  at {\it constant displacement steps}, rather than at constant {\it time steps}. This difference is crucial, because it is responsible for the computational advantage of the DRP approach.
In fact, the total Euclidean distance to be covered to transit from a coil configuration to the native state is typically only 1 order of magnitude larger than the most microscopic length scale, i.e. the typical monomer (or atom) size. As a consequence, only $30-100$ discretized displacement steps are generally sufficient for convergence~\cite{DFP1,DFP2}. This number should be confronted with the  $\mathcal{O}(10^{12})$ time discretization steps which would be required to accurately describe a $ms$-long folding transition, using uniform time steps, by sampling directly the integral (\ref{path2}) or by performing a long MD simulation. The physical reason for such an impressive simplification is that the DRP formalism avoids investing computational time to describe the time evolution of the system when it is not progressing towards the final state.

 Once the dominant pathways  have been determined, it is possible to systematically account for the effects of quadratic thermal fluctuations around them, by means of the Monte Carlo algorithm~\cite{DRP}.
 Let us denote by
\be
\ox(n)=({\bf \ox}_1(n),...,{\bf \ox}_{N_p}(n)),
\label{traj}
\ee
with $n=1,..,N$ a dominant  trajectory, i.e. a sequence of  molecule configurations determined by minimizing numerically the discretized HJ action (\ref{DSHJ}). 
The time at which each configuration $\ox(n)$ is visited during the transition can be obtained  by computing the set of time intervals separating each of the path steps $\ox(n)$ from $\ox(n+1)$:
\be
\Delta \tau_{n,n+1}= \frac{\Delta l_{n, n+1}}{\sqrt{4D (V_{eff}(\ox(n))-V_{eff}(\ox(1)))}}.
\ee

The sequence of time intervals obtained this way can be used to write a discretized version of the transition probability (\ref{path2}):
\be
P(x_f,x_i;\tau(N)) = \int~\prod_{n=1}^{N-1} \left[ d{\bf x}_1(n)... d{\bf x}_{N_p}(n)\right]~ e^{-\sum_{n=1}^{N-1} \Delta \tau_{n,n+1} ~\left[\frac{1}{4 D}~\sum_{i=1}^{N_p}~\left( \frac{{\bf x}_i(n+1)-{\bf x}_i(n)}{\Delta \tau_{n, n+1}}\right)^2 + V_{eff}({\bf x}_1(n), ..., {\bf x}_{N_p}(n))\right]}
\label{DPI}
\ee
We stress the fact that in Eq. (\ref{DPI}) the time intervals are chosen large (small), when the most probable trajectory at that time is evolving slowly (fast). 
In other words, the information encoded in the dominant reaction pathway has been used to identify a particularly convenient discretized representation of the path integral (\ref{path2}), in which the sizes of the time steps are adapted according to the evolution of the most probable pathways. 

Some comments on the discussion made so far are in order. First of all, it is important to emphasize that the saddle-point approximation underlying the DRP approach  is reliable only in the small temperature limit, in which the dominant pathways have little overlap and the contribution arising from the thermal fluctuations around them leads only to small corrections. Hence, an important question which will be addressed below is if this condition is satisfied  in the case of protein folding, at room temperature. A second observation is that, from the computational point of view,  the Monte Carlo sampling of the ensemble of fluctuations around each dominant reaction pathway in (\ref{DPI}) is essentially as expensive as finding the saddle-point path itself. In fact, in both cases one needs to sample a space with $3 N_p\times N$  degrees of freedom. 
Finally, we note that, in order to single out the next-to-leading corrections in the saddle-point approximation of the Fokker-Planck conditional probability, one should first expand the action to quadratic order around the dominant reaction pathways  and then perform the numerical  Monte Carlo sampling. However, in the regime of validity of the saddle-point approximation, this is not necessary. In fact,  the contribution to the Fokker-Planck conditional probability (\ref{path2})  of the cubic, quartic and higher terms lead to corrections which come in at next-to-next-to-leading order.  

\section{Dynamics of the Dominant Reaction Pathways}
\label{dDFP}

As discussed in section  \ref{inDRP}, the saddle-point paths satisfy the equations of motion associated to the effective action
\be
S_{eff}= \int_0^t d\tau \left(\sum_i^{N} \frac{\bf 1}{4 D}\dot{{\bf x}}_i^2\right) + V_{eff}({\bf x}_1,...,{\bf x}_{N}), 
\ee
where the effective potential $V_{eff}$ is defined in Eq. (\ref{Veff}). 

The exponent of the effective potential $e^{-V_{eff}(x) dt}$ represents the probability for the chain  to retain the same configuration $x$, during an infinitesimal time-interval $dt$. 
In particular, the first term in (\ref{Veff})  controls the tendency of the protein to rapidly evolve away from regions of configuration space where the forces are very large. We shall refer to this term as to the {\it force contribution} to the effective potential.
The second term in (\ref{Veff}) reduces the value of $V_{eff}$ in the vicinity of the narrow minima of the physical potential $U(x)$,  where the Laplacian is large and positive. This means that the chain will be trapped for long times in regions of configuration space characterized by a small conformational entropy. 
For this reason, we shall refer to such a term as  the {\it entropic} contribution to the effective potential $V_{eff}$. 

We note the fact that other approaches proposed in the literature to sample the relevant trajectories of Langevin dynamics have  ignored the entropic contribution to the effective potential  \cite{Doniach}. However, as it was  clearly shown in \cite{adib}, the inclusion of the entropic term is crucial in order to correctly identify the saddle-point pathways. 
In this section, we take a closer look to the role played by the entropic term in the specific case of molecular interaction. In particular, we estimate its  magnitude, relative to the force term, at room temperature.

Let us consider the effective potential associated to the interaction (\ref{Ugo}), between a single pair of residues.
We begin by observing that the entropic contribution coming from the harmonic interaction between consecutive residues on the chain reduces to an irrelevant constant, and therefore can be dropped. One the other hand,  the non-bonded interaction in (\ref{Ugo}) one has 
\be
v_{eff}(r) &=& \frac{D}{4 (k_B T)^2}\,\left[F^2 - 2 k_B T\, v_L\right]\\
F &=& 24 \,\epsilon \,\frac{\sigma^{6}}{r^{7}}\, \left[1- 2 \frac{\sigma^{6}}{r^{6}}\right]\\
v_L &=& 4 \,\epsilon \frac{\sigma^{6}}{r^{8}}\, \left[ 156 \frac{\sigma^{6}}{r^{6}}
-42\right].
\label{v2}
\ee

\begin{figure}
		\includegraphics[clip=,width=8 cm]{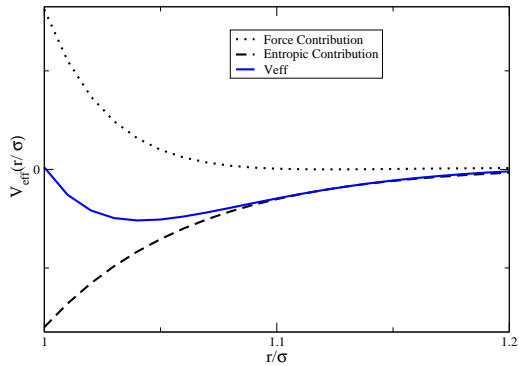}
	\caption{Comparison between the force and entropic contribution to the effective potential for a non-bonded
	Van-der-Waals type of interaction, Eq.(\ref{v2}).}
\label{compareForce}
\end{figure}

The numerical value of the force and entropic contributions to $v_{eff}(r)$ for different relative distances $r$ are presented and compared in Fig.\ref{compareForce}.
We note that, at distances smaller than the Van-der-Waals radius, the effective potential is dominated by the force contribution. This is expected, as  configurations in which two residues have some overlap are highly unstable under Langevin dynamics, because the effect of the force drift is very large.
On the other hand, it is interesting to observe that, outside the hardcore, the effective potential is dominated by the entropic contribution. 
In particular, entropic effects are responsible for a minimum of the effective potential at distances of the order of the Van-der-Waals radius.
Also this fact is easily interpreted: the configurations in which two monomers get close to each other without overlapping have a relatively long residence time, in Langevin dynamics. In fact, the short-range repulsion prohibits the system to evolve into configurations in which the two monomers come closer, while the Van-der-Waals attraction disfavors the escape toward configurations in which the monomers are further separated.  

From this discussion it follows that the stochastic dynamics of the polypeptide chain is the result of a delicate competition of entropic and energetic effects.  Approaches in which one retains only the force term  miss this important feature of the stochastic dynamics generated by the Langevin equation.

\section{DRP Calculation}
\label{calculations}

The DRP method yields the important reaction pathways connecting two well-defined system configurations, which have to be provided independently. Hence, the first step to  set up a DRP calculation consists in choosing the boundary conditions of the conditional Fokker-Planck probability (\ref{path2}).  
In many reactions of interest, such as  cis-trans isomerization or allosteric transitions, both the initial and final configurations are well-defined  and can be  experimentally determined.
However, in the specific context of the study of the protein folding reaction, the denatured conformations populate a huge portion of the configuration space and cannot be determined experimentally.
Under such  conditions, one has to average over the transitions  between a representative set of
different denatured conformations and the native state.
The DRP retains its computational usefulness only if the folding transition can be accurately characterized by averaging over a relatively small number of denatured configurations
We generated the set of denatured configurations by performing $\sim10^4$ steps of MD simulation at $500 K$, starting from the native state, followed by few thermalized MD steps at $300 K$. This way, we observed a relatively rapid unfolding of the chain. 

The next step in DRP calculations consists in finding a representative set of dominant pathways,  by minimizing numerically the HJ action (\ref{SHJ}), for each of the boundary conditions which have been previously determined.
Any algorithm for the minimization of the HJ functional has to be provided with a starting path connecting the initial and final points, from which it begins  the search for the most probable trajectories.  
The trajectories used to generate the denatured configuration by  high-temperature MD simulations represent a natural 
and convenient choice, which has the advantage to avoid residue overlap, covalent bond breaking, etc... However, such MD trajectories contain a very large number of discretization steps $N_{MD}~\sim~10^4$, while the main advantage of the DRP method is that it allows to keep only a small number of path steps $N_{drp}$, typically of the order of $30-100$. Consequently, before beginning the minimization procedure, one needs to reduce the number of path steps in the initial MD trajectory from $N_{MD}$  down to $N_{drp}\ll\,N_{MD}$. 
This was done by applying the iterative smearing algorithm introduced in \cite{DRP}:  at each iteration, we defined a new path $\{{\bf x}'(i)\}_{i=1...N_{MD}}$, consisting of $N_{MD}/2$ steps, in which each configuration was defined as average of two configurations of the input path at consecutive time steps, i.e. 
\be
{\bf x}'(i)=\frac{1}{2}\left({\bf x}(2 i +1)+{\bf x}(2 i) \right).
\ee
After few of such decimation steps, we obtained a smeared periodic trajectory consisting of $N_{drp}=50$ steps.

Occasionally, the averaging involved in the smearing algorithm led to  paths which contained configurations in which the hardcore of some of the residues overlap. Under such circumstances, we observed a huge increase of the HJ effective action and the rejection rate of the relaxation algorithm was very large. In order to accelerate the convergence of the optimization algorithm,  we ran a short preliminary energy minimization of each of the configurations in the starting path.

The relaxation of the HJ action was performed via a simple adaptive simulated annealing algorithm. After a preliminary thermalization phase, based on the Metropolis algorithm, we performed 100 cooling cycles, consisting of 10 steps each. Each move was obtained by up-dating the path globally, i.e. moving the positions of all the particles, in each configuration of the path.  At the end of each cooling cycle, the boldness of the moves was adjusted, in order to keep the rejection rate between $50\%$ and $90\%$.

After generating the set of dominant reaction pathways, we sampled the ensemble of thermal fluctuations around each of them according to the algorithm introduced in \cite{DRP} and illustrated in section \ref{inDRP}, i.e. by sampling the
path integral (\ref{DPI}) through a Monte Carlo simulation, using the saddle-point paths determined before as the starting points of the Markov chains. We adopted the Metropolis algorithm and sampled 10 thermal fluctuations for each dominant reaction trajectory.
\begin{figure}
		\includegraphics[width=8 cm]{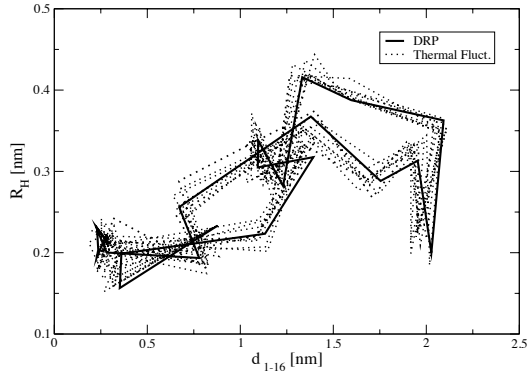}\quad
			\caption{Structure of the thermal fluctuations around a dominant reaction pathways.  The solid line represents the local dominant reaction pathways, the dotted lines represent 10 independent thermal fluctuations around the dominant reaction pathways.}
\label{fluct}
\end{figure}

\section{Results and Discussion}
\label{results}

In this section we analyze the results obtained with the DRP method. We study the ampitude of thermal fluctuations around the dominant reaction pathways obtained at room temperature, and we compare the outcome of the DRP analysis with the results obtained from MD simulations.

\subsection{The consistency of the saddle-point approximation}

The first question we address is whether the saddle-point approximation underlying the DRP approach is consistent, when applied to the study of the chain's conformational transitions, in the present model. 
A first necessary condition is that, at room temperature $T=300 K$,  thermal fluctuations around the dominant pathways must be  small. More precisely, including quadratic fluctuations  should provide only small corrections to the leading-order predictions. 
A second necessary consistency condition is that thermal fluctuations around different saddle-points must  not significantly overlap\footnote{Note that these  conditions are the analog of those required  for the validity  of the semi-classical approximation in Quantum Mechanics.} \cite{DRP}. 

In order to verify if the first of such conditions is fulfilled, in Fig. \ref{fluct} we compare different types of trajectories, in the plane selected by the $R_H$ and $d_{1-16}$  reaction coordinates. The solid line represents the result of the minimization of the HJ action, i.e.  a  dominant reaction pathway. The dotted lines are 10 independent thermal fluctuations obtained by  2000 steps of Monte Carlo sampling of the path integral (\ref{DPI}), starting from the same dominant trajectory.
We observe that the thermal fluctuations shown in Fig.\ref{fluct}, remain close to and qualitatively similar to the nearby saddle-point path. This feature is not altered when one considers longer Markov chains in the Monte Carlo sampling of the thermal fluctuations, via the path integral (\ref{DPI}). 

Let us now discuss the second consistency condition, i.e. whether thermal fluctuations associated to different dominant reaction pathways  significantly overlap. If this was the case,  then we would expect  that the relaxation of the HJ action performed  starting from fluctuations associated to a given saddle-point  would occasionally converge to a different saddle-point. 
However, for all the 8 dominant  pathways considered, the annealing of the HJ action performed starting form  each of the corresponding thermal fluctuations converged very rapidly to the  original saddle-point trajectory. This fact gives us some confidence that the thermal fluctuations associated to different minima of the HJ action do not significantly overlap and that the saddle-point expansion underlying the DRP method is working well.  

\begin{figure}
		\includegraphics[angle=-90,width=6.8 cm]{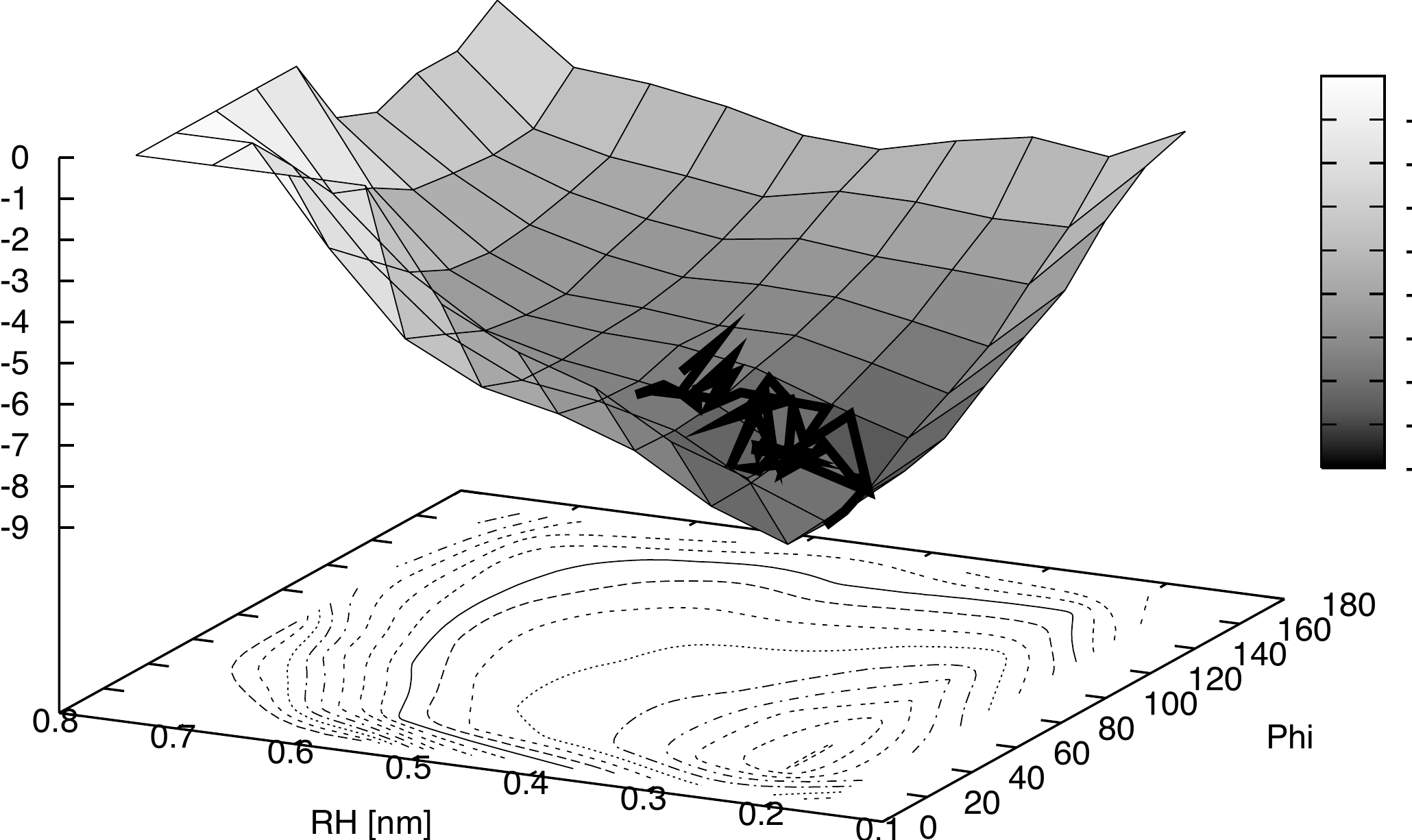} 
		\hspace{1.8 cm}
		\includegraphics[angle=-90,width=7.7 cm]{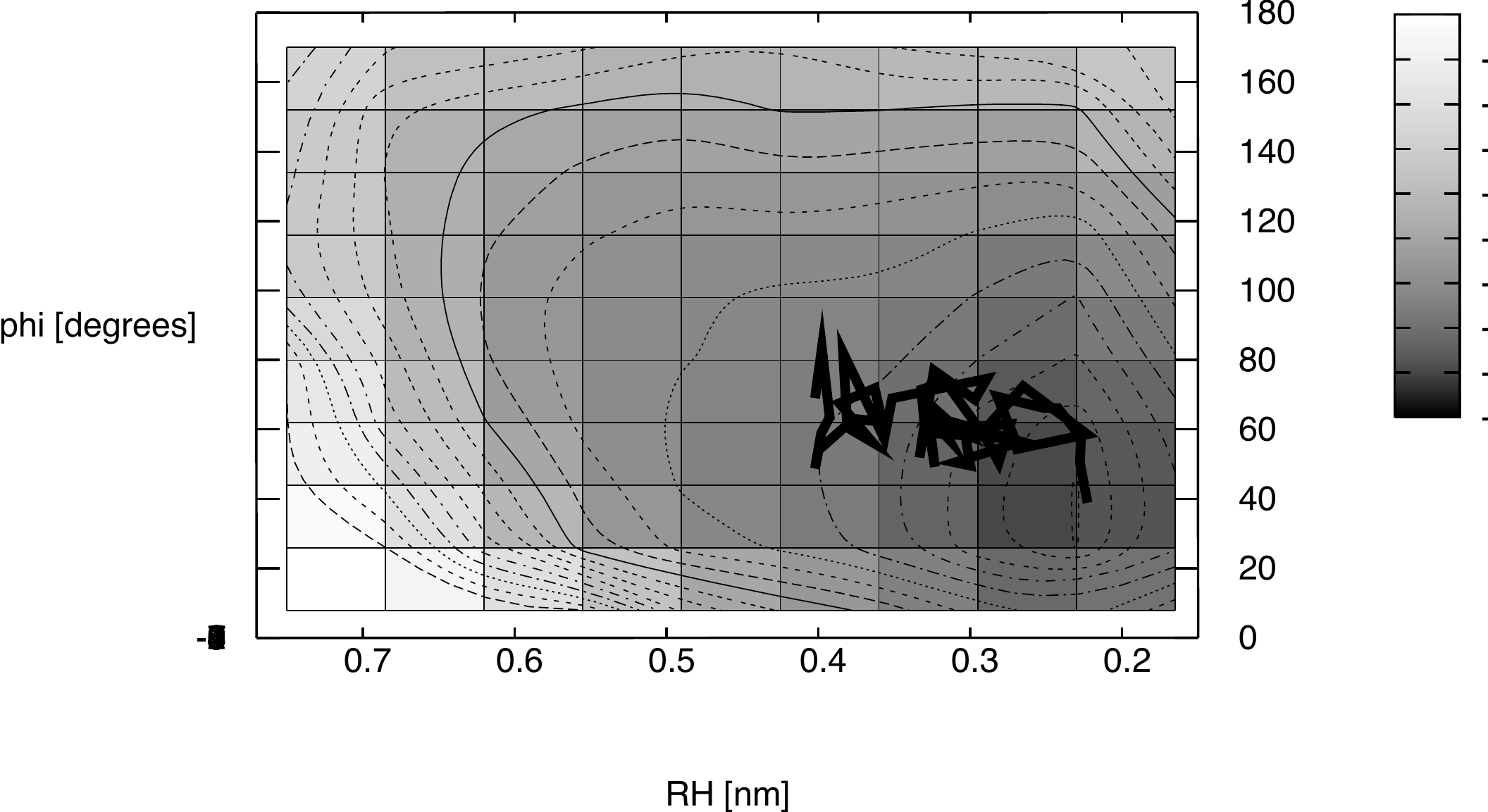}\\
		 \vspace{1 cm}
		\includegraphics[angle=-90,width=6.8 cm]{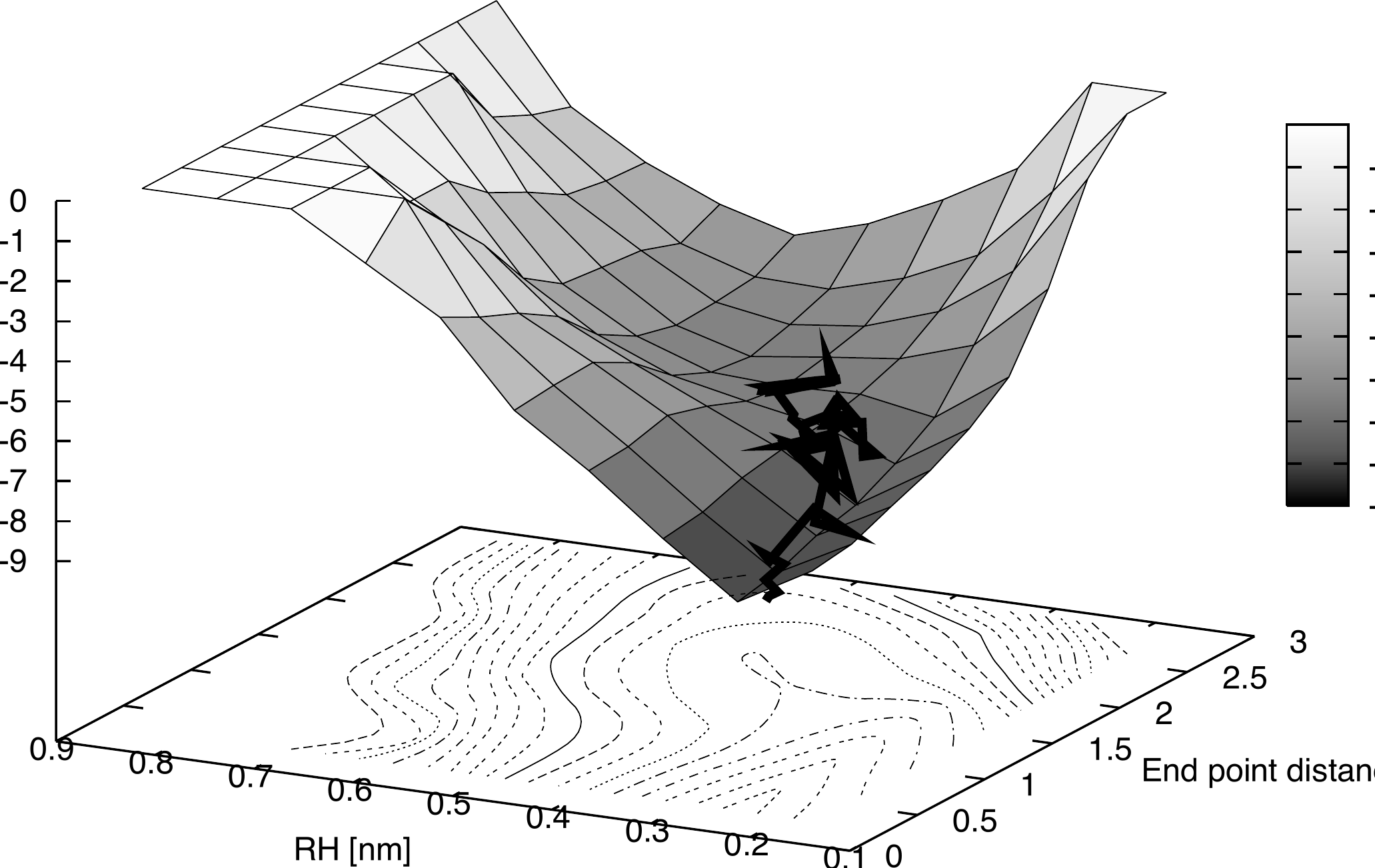}
		\hspace{2 cm}
		\includegraphics[angle=-90,width=7.6cm]{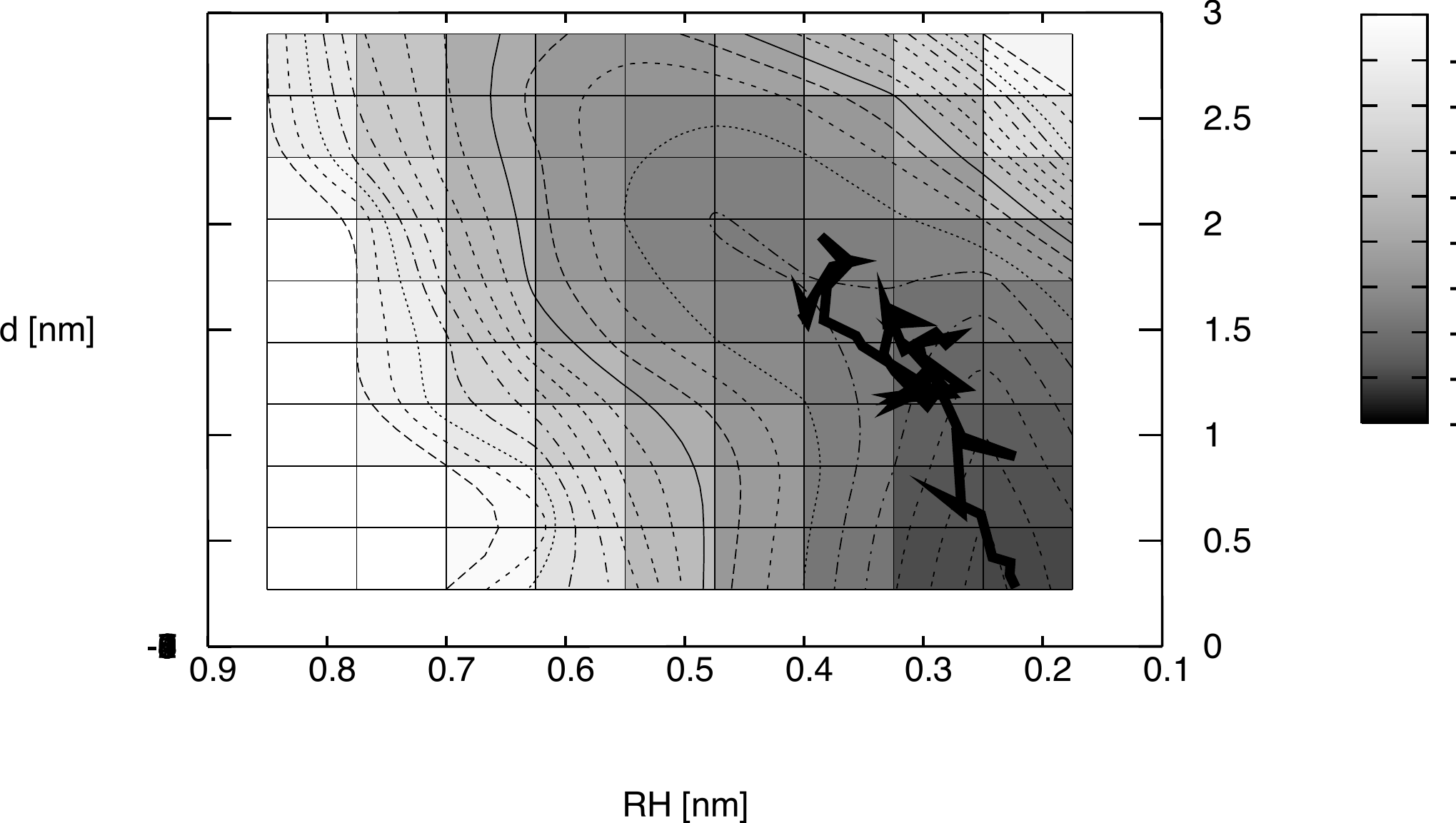}\\ 
		\vspace{1 cm}
		\includegraphics[angle=-90,width=6.8 cm]{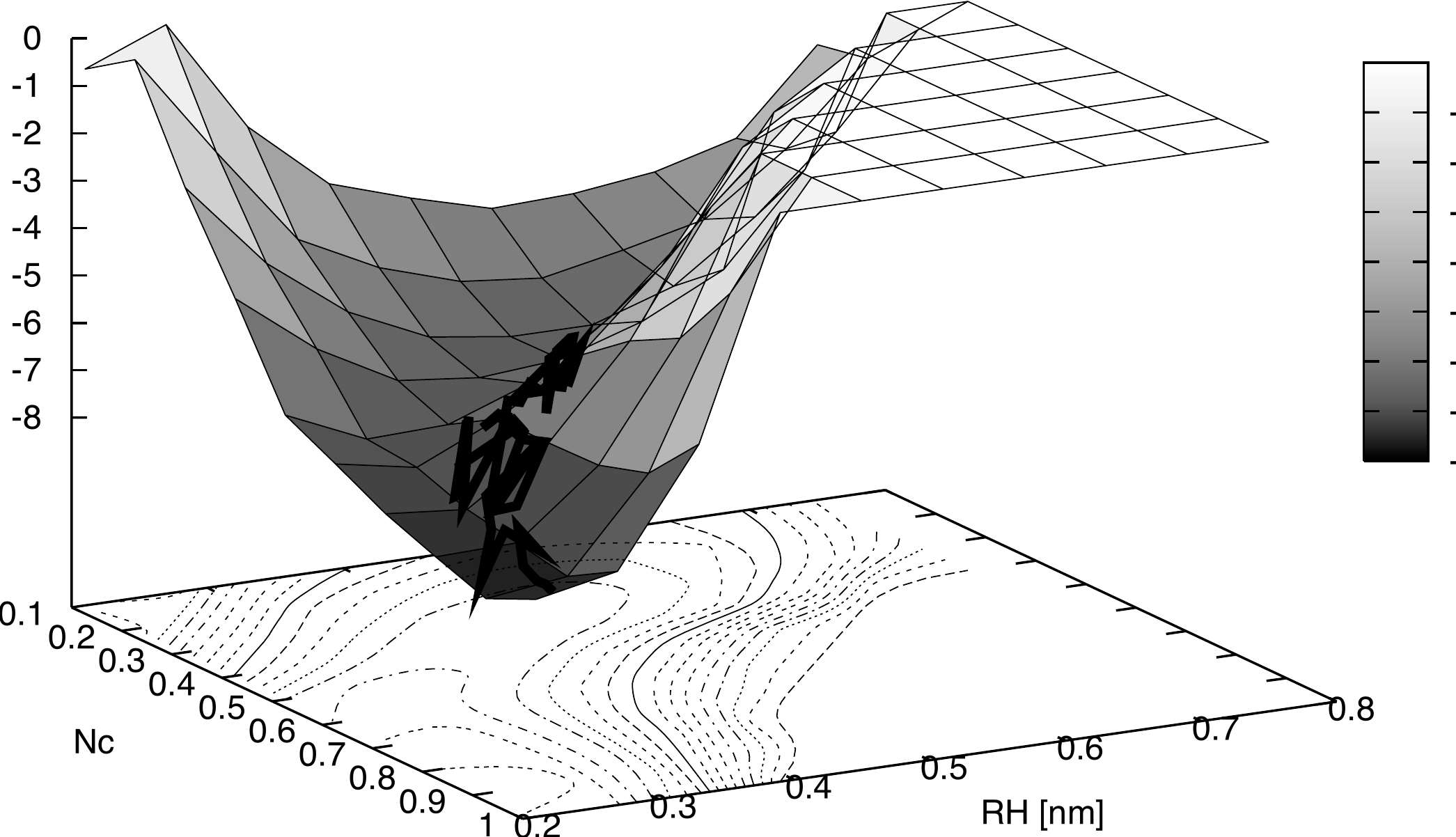}
		\hspace{2 cm}
		\includegraphics[angle=-90,width=7.5 cm]{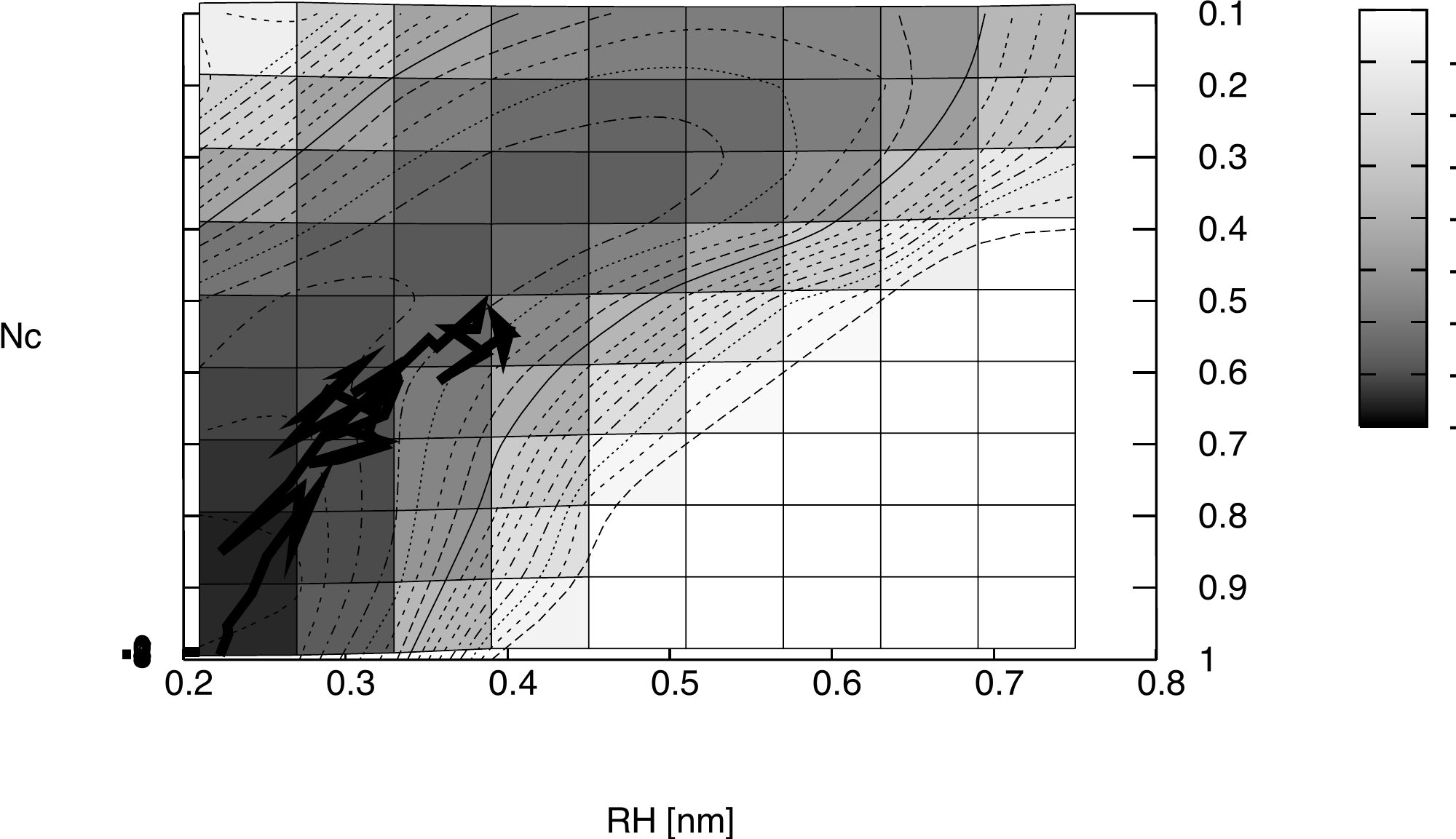}
	\caption{Comparison between MD and DRP calculations for three pairs of order parameters ($d_{1-16}$ vs $R_H$, $R_H$ vs $\phi$ and $R_H$ vs $N_c$. The free energy landscape has been obtained by MD simulations, while the line represents the trajectory obtained averaging over 8 dominant reaction pathways..}
\label{G3}
\end{figure}

\subsection{Comparison with MD}

A  fundamental condition for the DRP analysis to be useful is that it is possible to reliably identify the regions of configurations space which are most probably visited during the folding reaction by averaging over {\it a moderate number} of independent dominant reaction pathways trajectories. 
To check if this is the case for the present model, in Fig.\ref{G3} we compare the results of DRP simulations with the free energy landscapes obtained by long MD trajectories, in three surfaces selected by the  order parameters $d_{1-16}$, $R_H$, $N_c$ and $\phi$.
The line represents the most probable trajectory, which was obtained averaging over only 8 independent dominant reaction pathways. In total, this calculation required about 6 CPU hours. 

We can see that the DRP method correctly identifies the regions of configuration space which are most likely explored by the MD trajectories. In fact, the line obtained averaging over the 8 initial configurations travels along the valley in the free energy profile until it approaches a region where the free energy is nearly flat. It should be stressed that the DRP calculation does not need to make any assumption about the structure of the free energy, nor it requires any choice of reaction coordinate.  The trajectories obtained from averaging over the dominant reaction pathways are parameter free predictions and the agreement with MD results is observed in all the combinations of the selected set of reaction coordinates.

\begin{figure}
		\includegraphics[width=5.7 cm] {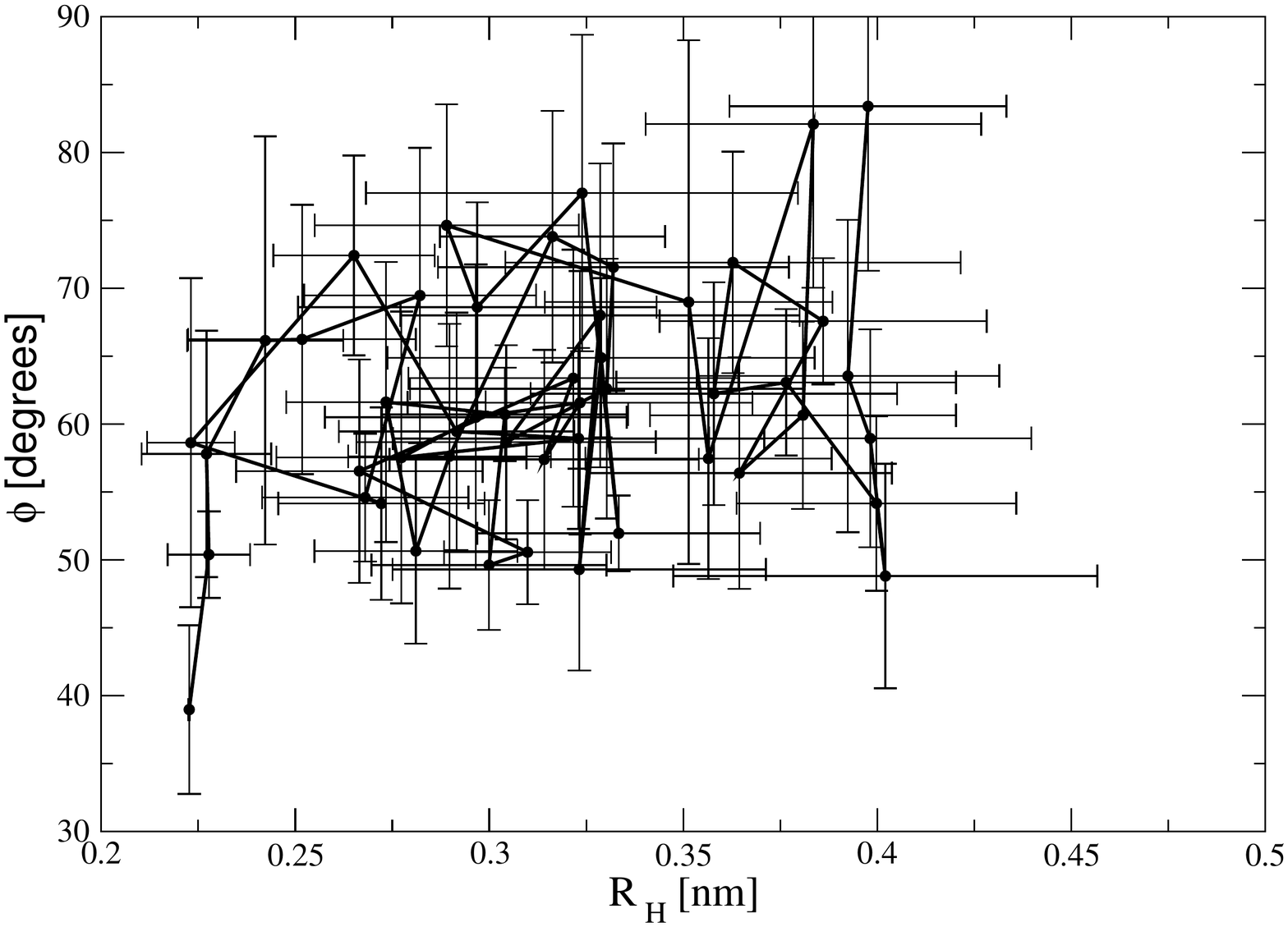}\hspace{0.2cm}
		\includegraphics[width=5.7 cm] {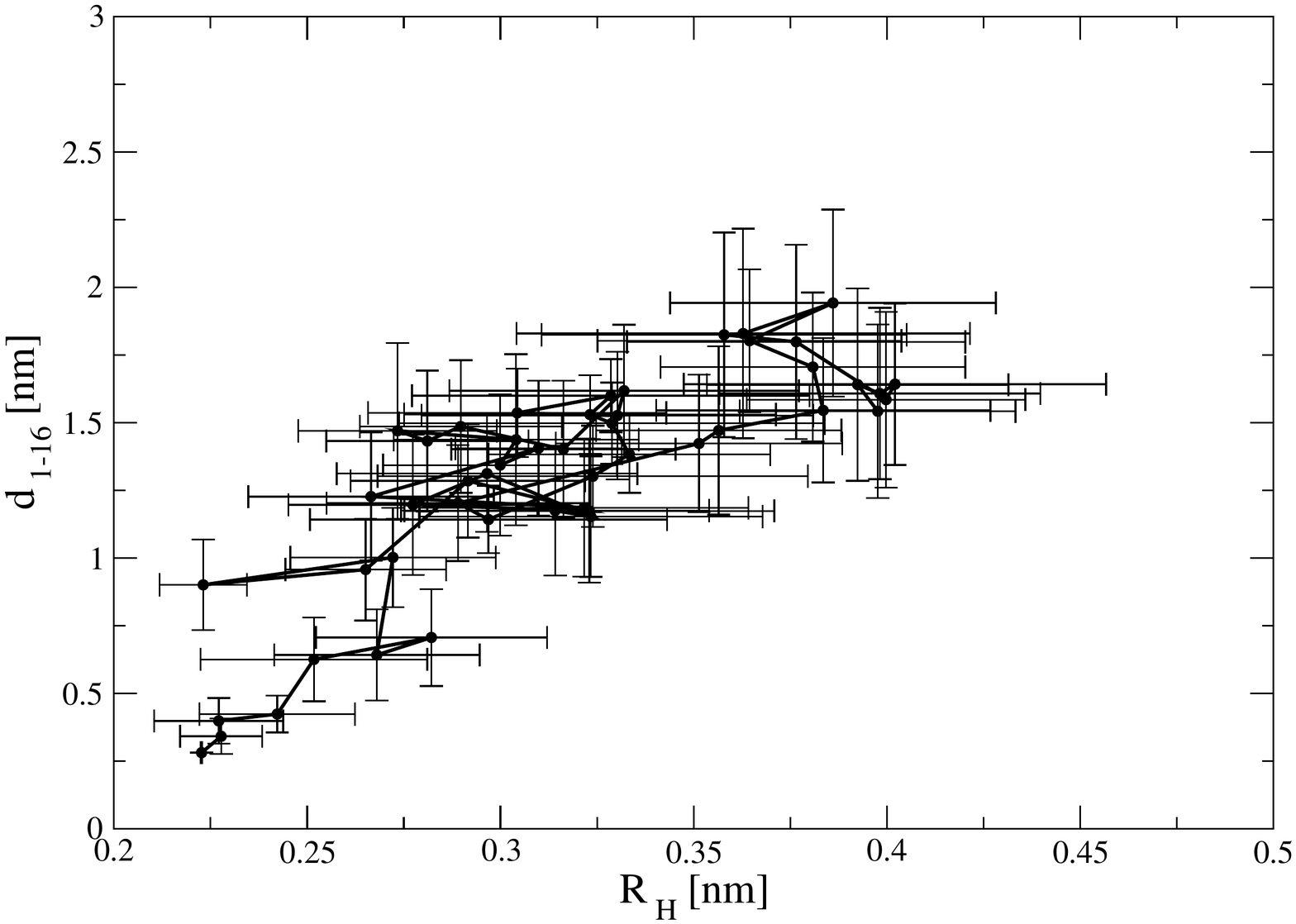}\hspace{0.2cm}
		\includegraphics[width=5.7  cm] {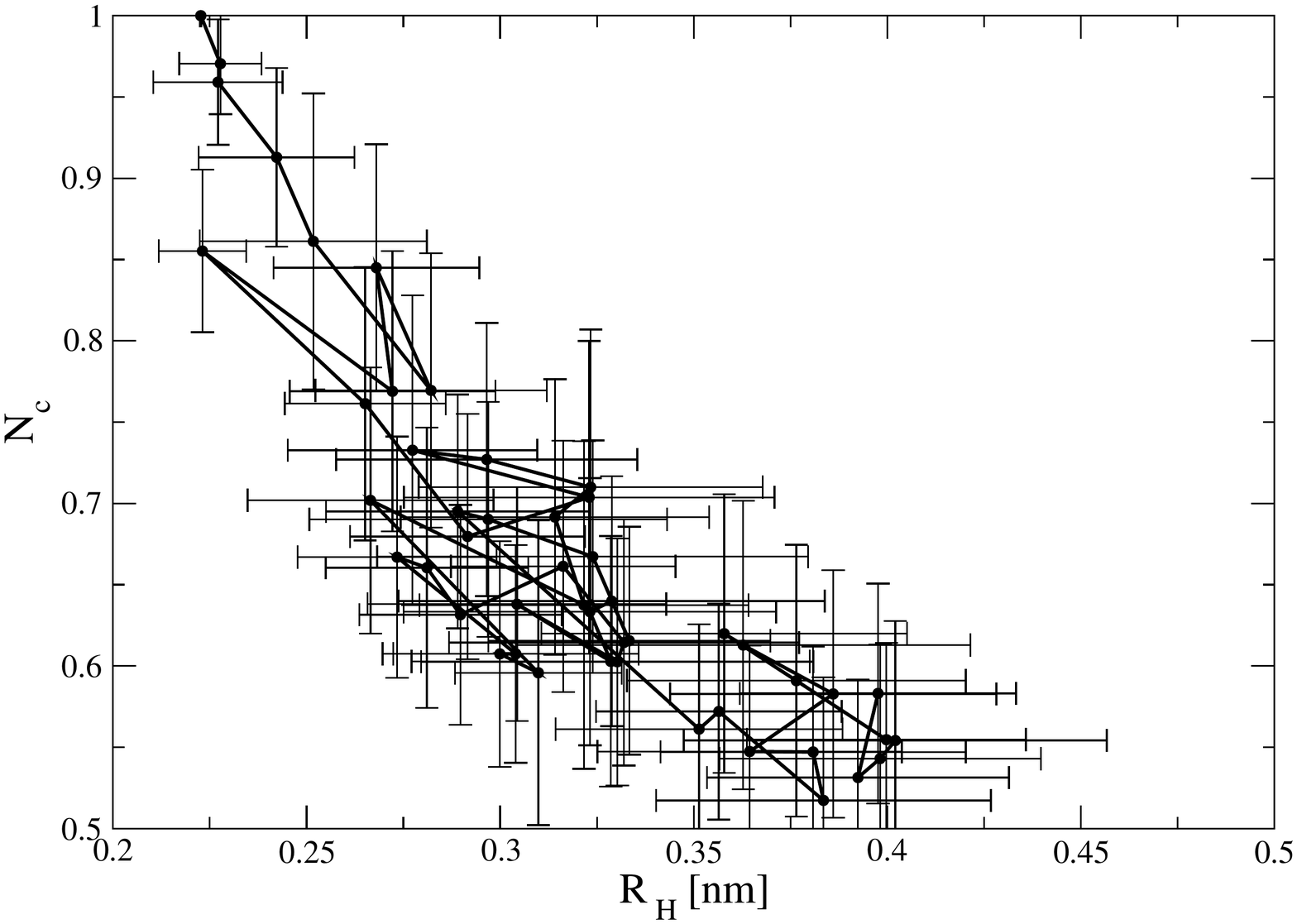}
	\caption{Statistical errors on the trajectory obtained by averaging over 8 dominant reaction pathways in  three surfaces selected by the parameters $d_{1-16}$, $N_c$ and $\phi$.}
\label{G3e}
\end{figure}

An important related question to address is what precision can be achieved by averaging over a moderate number of initial configurations. 
To answer, in Fig.\ref{G3e} we plot the error on the average,  obtained from the standard deviation calculated from the 8  trajectories. Even with such a small number of trajectories, the statistical error on the average  is rather small, compared with the scale at which the free energy profile varies significantly. 
  
We stress the fact that the number of dominant  pathways considered is almost three orders of magnitude smaller than the number of unfolding/refolding transitions which were required to obtain the free energy landscape from  MD simulations. 

\begin{figure}[b]
		\includegraphics[width=5.7 cm]{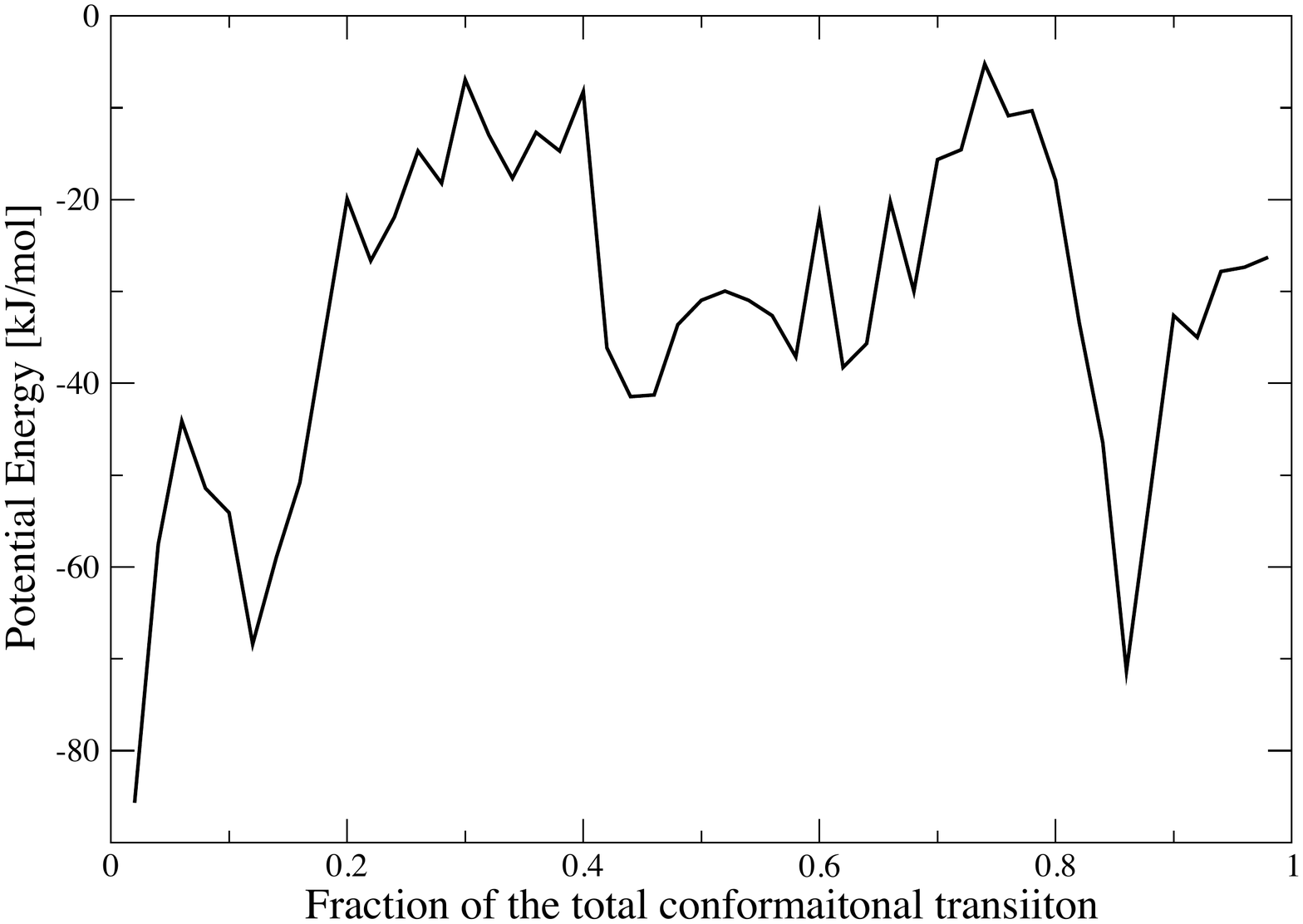} \hspace{0.2cm}
		\includegraphics[width=5.7 cm]{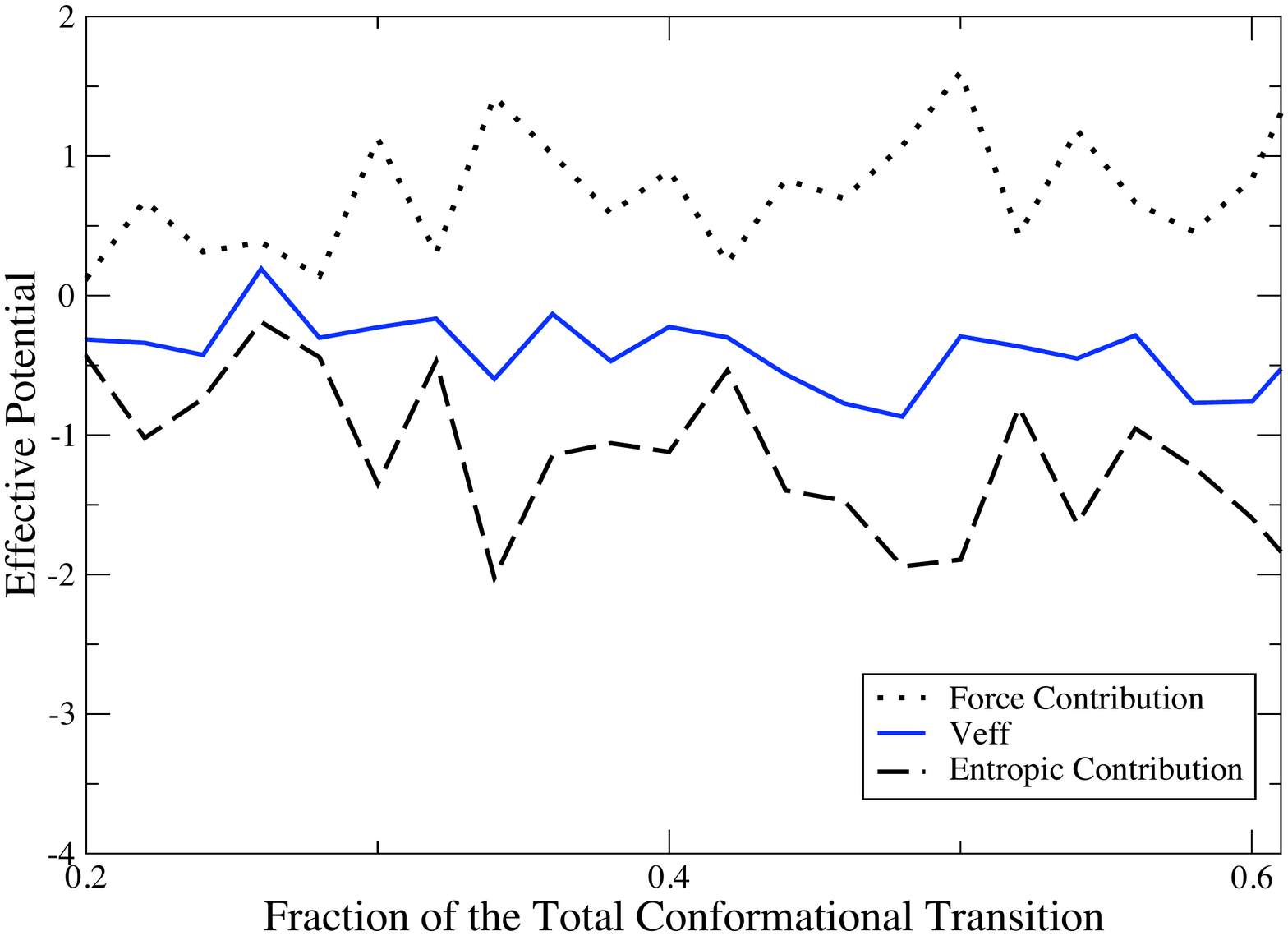}\hspace{0.2cm}
		\includegraphics[width=5.7 cm]{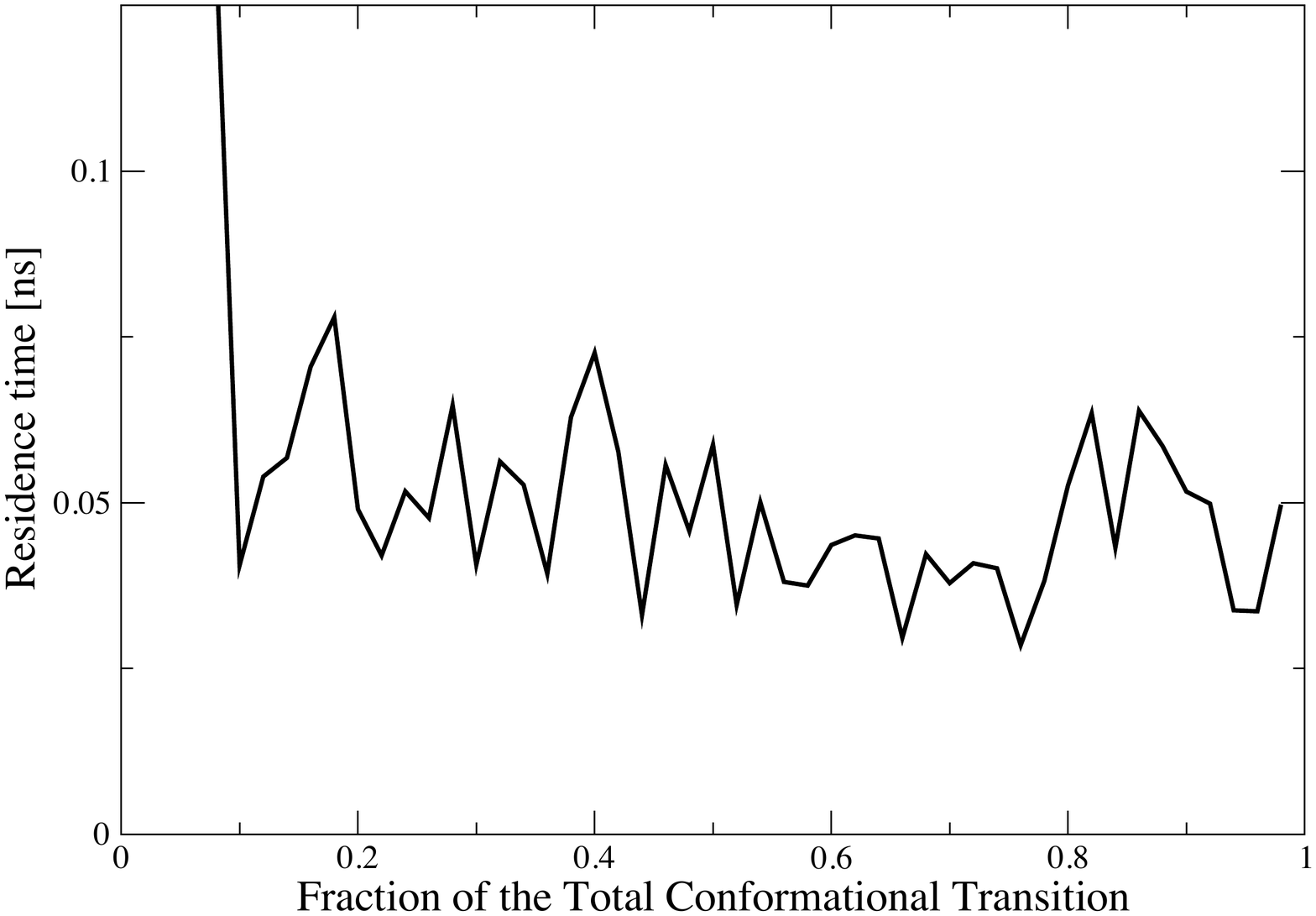}
	\caption{Evolution of the total potential energy $U(x)$ (left panel), effective potential $V_{eff}(x)$ (center panel) and residence time (right panel) along a typical dominant reaction pathways.}
\label{resDFP}
\end{figure}

In Fig.\ref{resDFP} we plot the potential energy $U(x)$, the effective potential $V_{eff}(x)$ and  the residence time along a typical path.
From the left panel, we can see that the transition involves overcoming several energetic barriers. The fact that the DRP method finds the optimal path in the free energy landscape while paying a price in energy suggests that the contribution of entropic effects is significant. This fact is confirmed by the plot presented in the center panel of Fig.\ref{resDFP} , where the  {\it force} and {\it entropic} contributions to the effective potential are compared and found to be of the same order (see discussion in section \ref{dDFP} and in \cite{adib}).  The relaxation of the HJ action drifts the paths towards regions of configurations space  where the forces are small and the Laplacian is large and positive.  The third panel of Fig. \ref{resDFP} displays the residence time along a typical dominant trajectory.  We see that the residence time diverges in the vicinity of the native state, while the molecule spends very little time in each of the denatured conformation. This is  what one expects from general arguments. In fact, equilibrium experiments indicate that at room temperature,typically  about $80\%$ of the chains are in the (almost unique) native conformation, while the remaining $20\%$ are in denatured conformations.  Since the conformational entropy of the denatured states is huge, the time spent by the chain in each denatured conformation must be exponentially small.
 
\section{Folding Mechanism for the $\beta$-Hairpin}
\label{mechanism}

Although the focus of the present work is  on assessing the reliability  of the DRP method, it is instructive to discuss the folding mechanism suggested by  the present simple Go-type model. This section can be considered as preparation work for a forthcoming analysis, which will be based on atomistic force fields, as it was done in \cite{DFP2} for alanine dipeptide.

The problem of identifying the mechanism responsible for the folding of a $\beta$-hairpin has been the subject of an intense theoretical and experimental investigation (see e.g. \cite{Eaton1} and references therein). 
Eaton and co-workers have proposed a picture in which the stabilizing hydrogen bonds on the backbone are formed one by one, and are initiated at the turn (so called "zipper" mechanism) \cite{Eaton1}.
On the other hand, several theoretical simulations performed in simplified models \cite{simple} and in all-atom models with implicit \cite{impl} and explicit \cite{expl, Doniach} solvent seem to favor a different folding mechanism, in which after the initial $\beta$ turn, the hydrophobic residues collapse into a hydrophobic core, which is then followed by stabilizing bond formations.

The two pictures lead to predictions which can be tested by measuring of the folding rates of mutated variants of the hairpin.  In the zipper mechanism, a mutation which increases the stability of the hydrophobic cluster is not expected to significantly alter the folding rate, since the contacts in the hydrophobic core are formed after the transition state. 
In the hydrophobic cluster mechanism, the folding rate should not depend on the type of residues in the turn region. In addition, assuming that cluster is partially formed at the transition state, this picture predicts that the folding rate should be larger for sequences containing  stabler hydrophobic clusters.

Experiments seem to be overall more consistent with the zipper mechanism. 
For example, measurements performed by Gai and co-workers \cite{Gai} have shown that the hairpin called {\it trpzip4} (sequence GEWTWDDATKTWTWTE), folds at half of the rate of GB1 although  has the same turn sequence of GB1, and is 15-fold more stable. This result is inconsistent with the hydrophobic collapse mechanism. 

There are several possible reasons for the observed discrepancy between theory and experiment. On the one hand,  the available force fields may not accurate enough to describe the long-time dynamics of a polypeptide chain. On the other hand, the algorithm used to generate folding trajectories from a given force field may be spoiled by wrong approximations or by the choice of some "bad" reaction coordinates. Finally, it is possible that the folding trajectories  used to infer the folding mechanism  do not form a statistically significant sample. 
The DRP method provides in principle a way to test directly the validity of a given force field. In fact, it yields by construction  the most statically significant  trajectories, without involving any {\it a priori} choice of reaction coordinate, and without relying on any ad-hoc assumption, other than the underlying Langevin Eq.
 
Let us begin our discussion by inferring  what mechanism is suggested by our MD simulations.
The lowest  panel of  Fig.\ref{G3} shows that the configurations which are visited when the MD trajectories leave (or return to ) the native state are characterized by an almost constant size of the hydrophobic cluster (between $0.2 nm$ and $0.3 nm$, while the fraction of native contacts  drops significantly (from 1 to 0.6). This feature is clearly suggestive of a hydrophobic collapse mechanism.

\begin{figure}
		\includegraphics[width=8 cm]{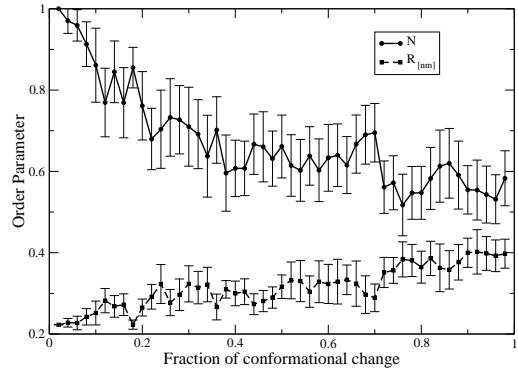}
	\caption{Evolution of the order parameters $R_H$ and of the fraction of native contacts $N_c$ as a function of the fraction of conformational changes, averaged over 10 independent dominant reaction pathways trajectories.}
\label{DFPall}
\end{figure}

\begin{figure}
\centering
\includegraphics[width=3.3 cm]{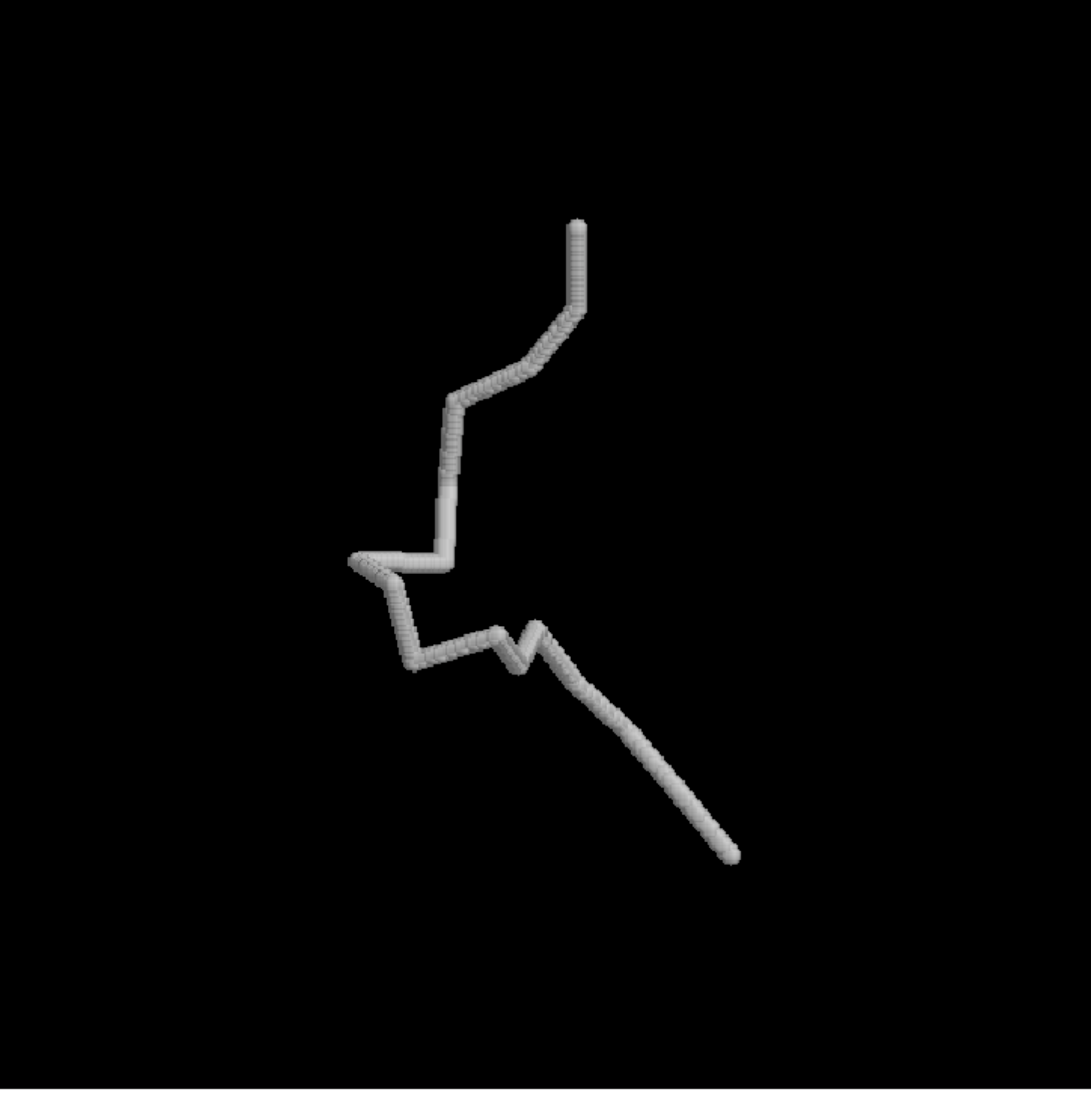}
\includegraphics[width=3.3 cm]{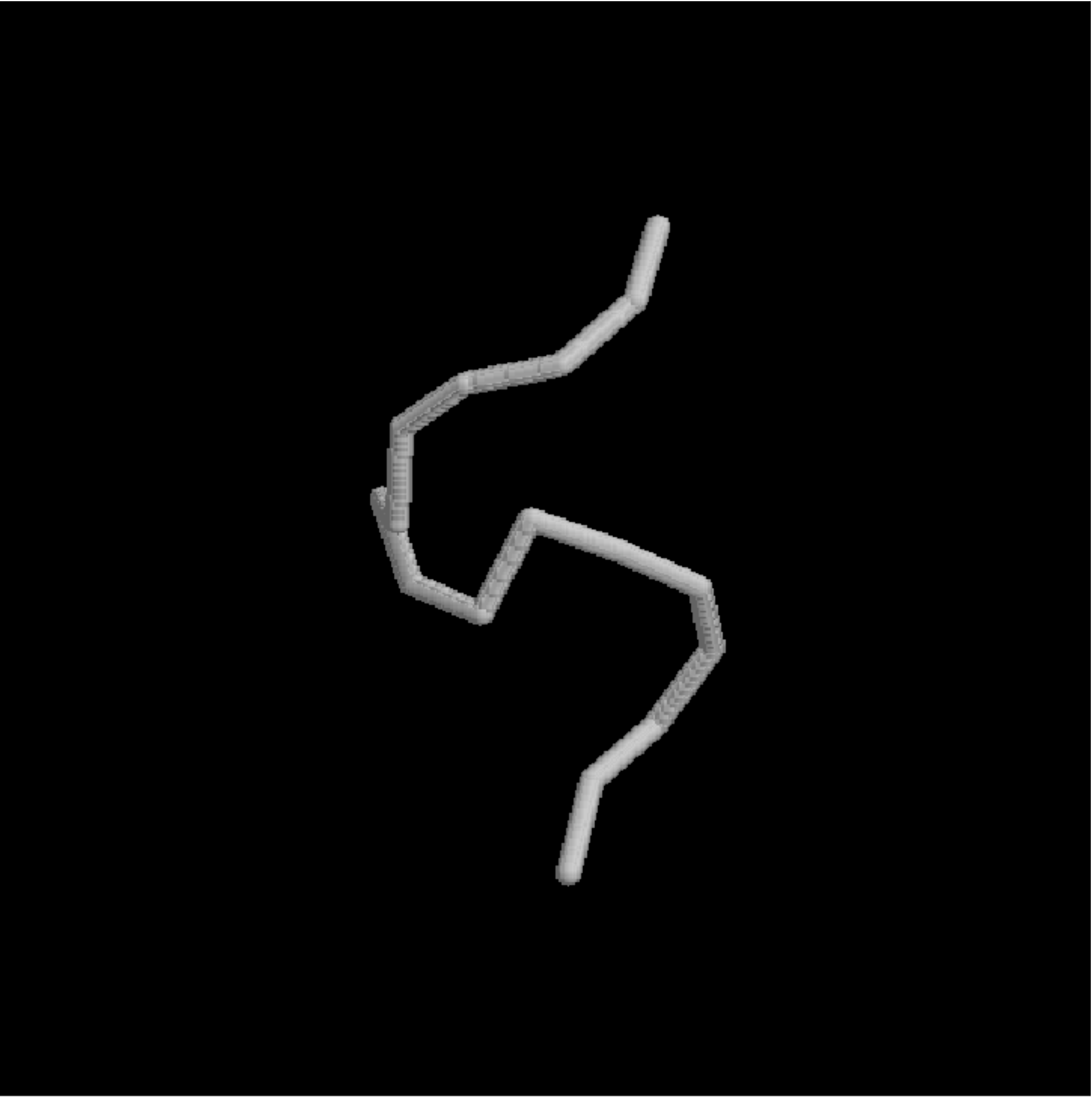}
\includegraphics[width=3.3 cm]{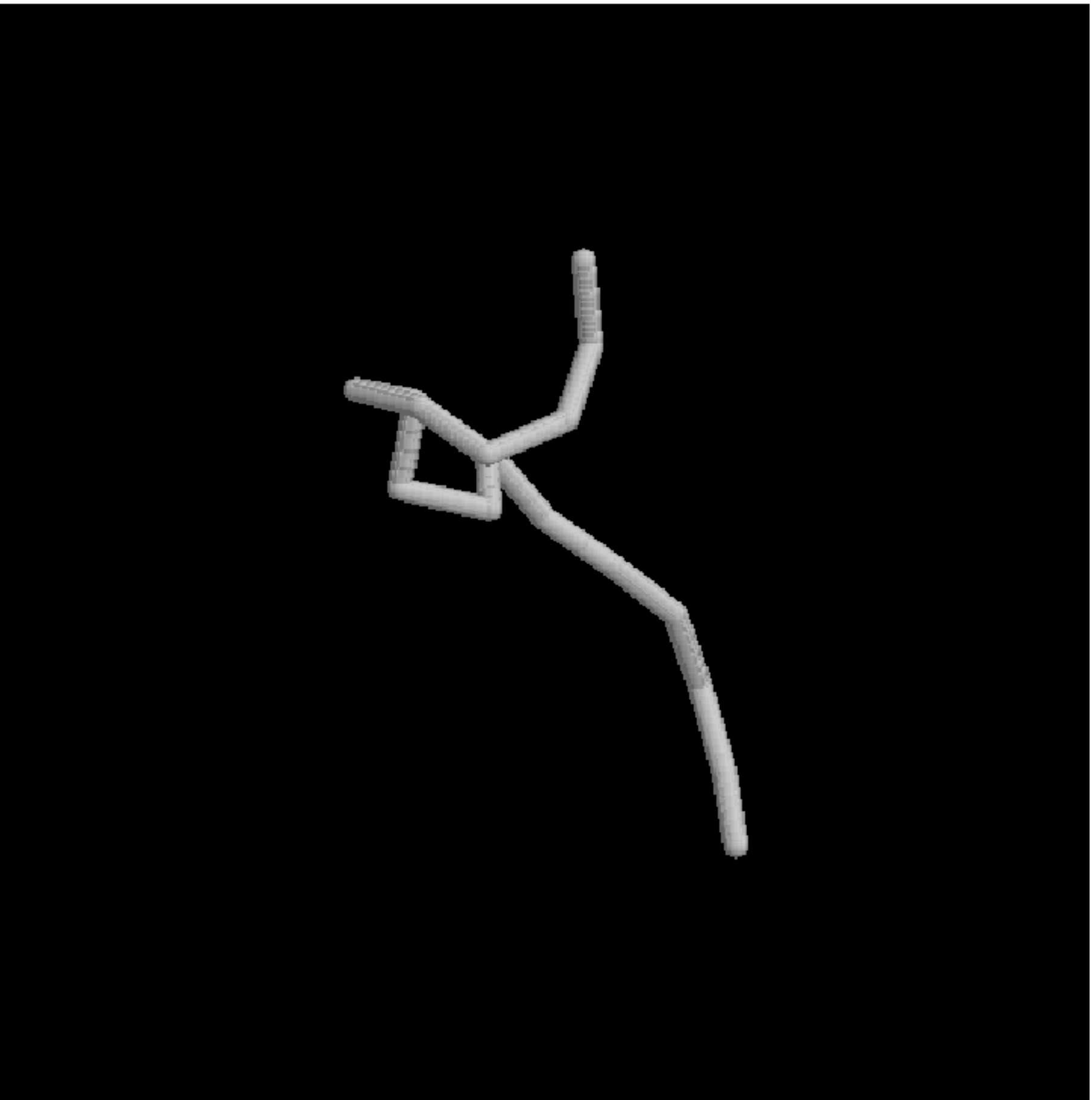}
\includegraphics[width=3.3 cm]{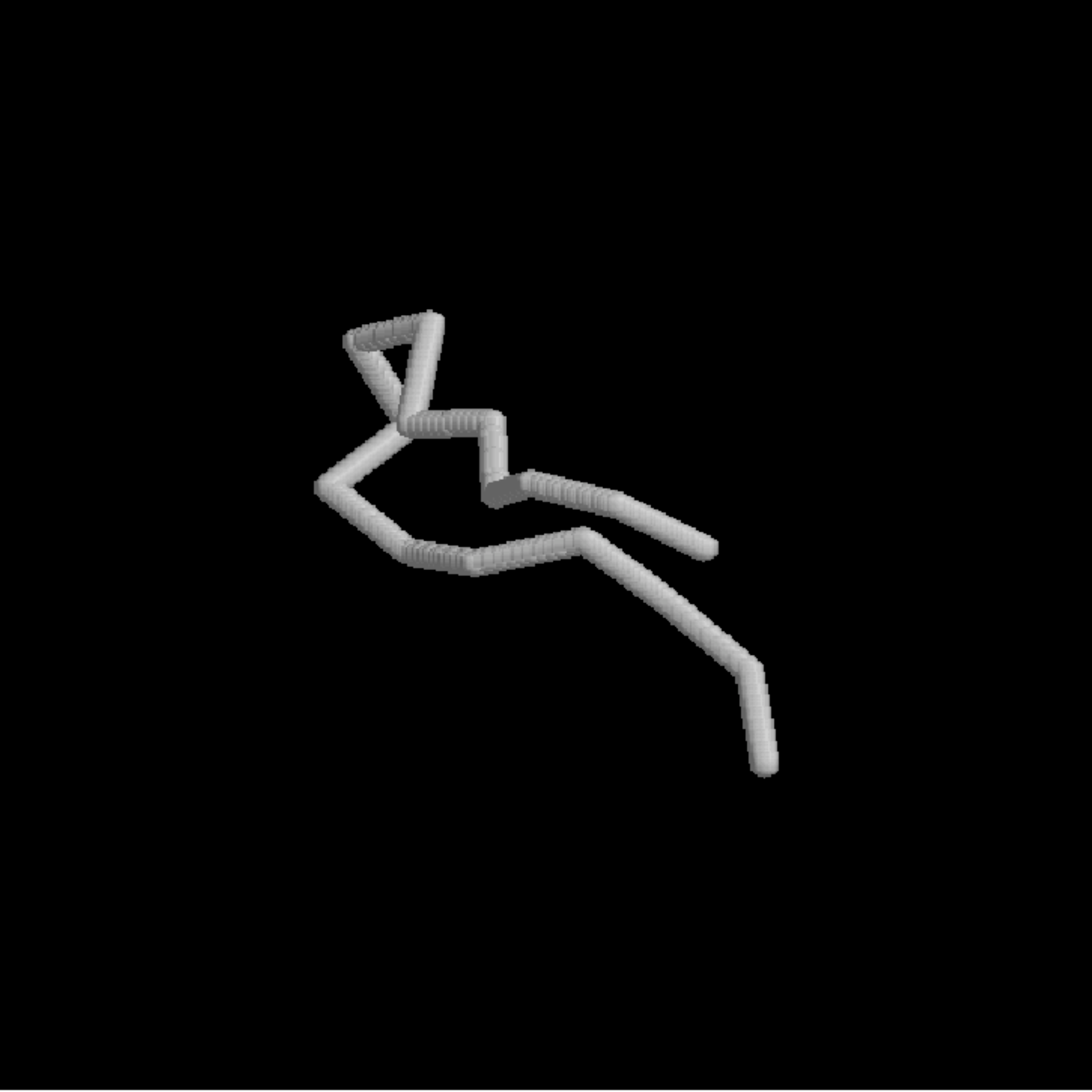}
\includegraphics[width=3.3 cm]{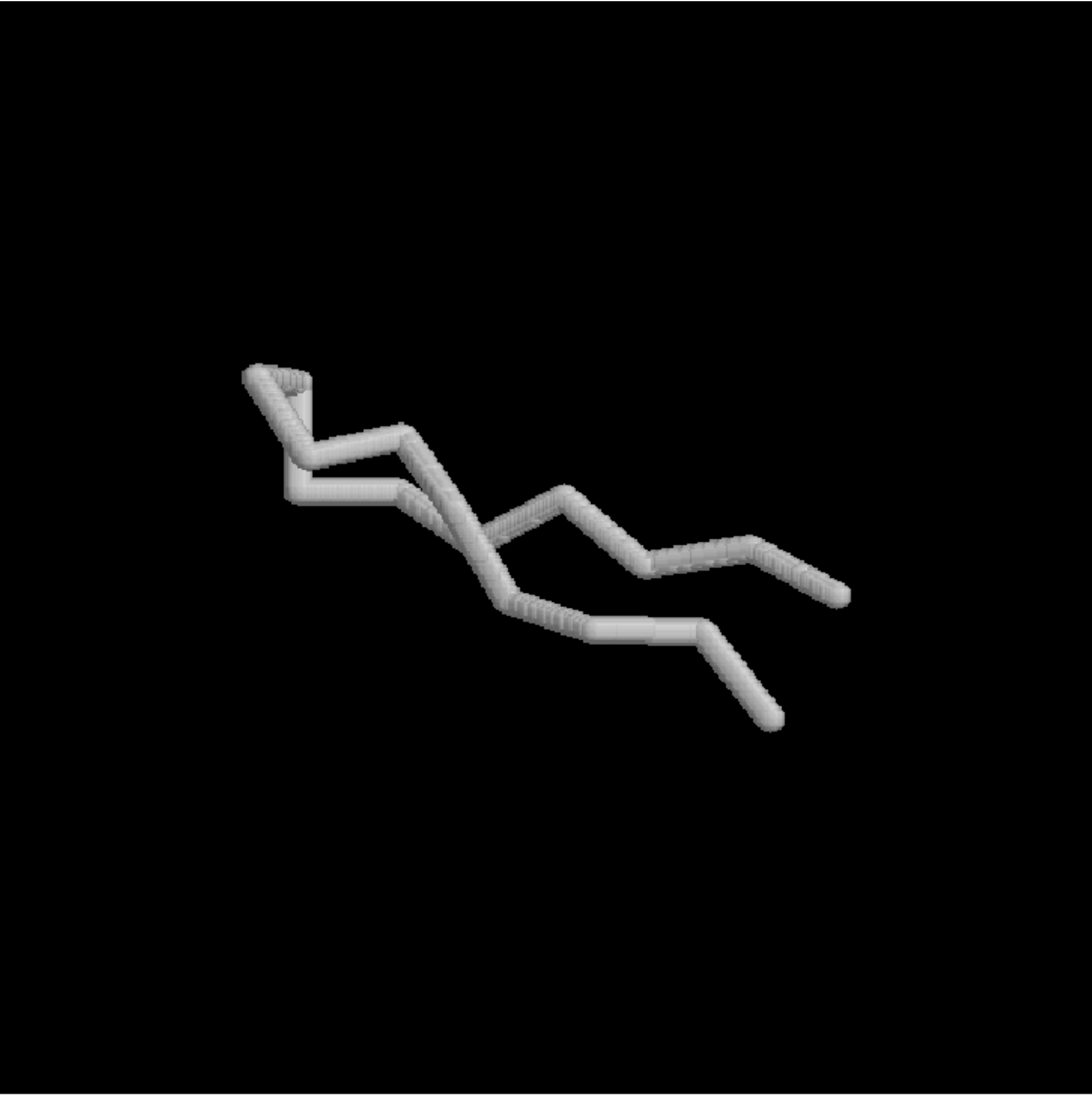}
\caption{Sequence of the hairpin configurations along the dominant reaction pathways. This sequence illustrates how the folding is initiated by the formation of the contact between the hydrophobic residues.}	\label{fig:10001}
\label{frames}
\end{figure}

The same conclusion could have been reached directly from the DRP calculation, without having to generate the free energy landscape by MD. 
In Fig.\ref{DFPall} we compare the average evolution of the gyration radius of the residues in the hydro-phobic cluster $R_H$ with the fraction of native contacts $N_c$, as a function of the fraction of the  total number of path steps \footnote{It should be stressed that the equations defining the dominant reaction pathways are invariant under time-reversal, therefore the folding and unfolding transitions are completely equivalent, in this formalism.}. Again, the average was performed over just 8 trajectories.
We observe that, during the first $20\%$ of the unfolding transition, the gyration radius of the hydrophobic core remains statistically consistent with its value in the native conformation. On the other hand, in the same part of the transition, the fraction of native constants  drops by about $40\%$, in quantitative agreement with MD results.  An example of hydrophobic collapse observed in a dominant transition is provided by the sequence of  conformations presented in Fig. \ref{frames}.

\section{Conclusions}
\label{conclusions}
In this paper, we have assessed the reliability of the DRP approach,  when it is applied to study  protein folding reactions. As a test problem, we studied the folding of 16-residue $\beta$-hairpin, using a Go-type reduced model. 

First, we have provided evidence that the saddle-point approximation ---which underlies the DRP approach--- is working well, at room temperature. In fact, we have verified that the thermal fluctuations around the different dominant reaction pathways  are small and do not destroy the picture provided by the lowest-order analysis. In addition, we have shown that fluctuations around different saddle-point paths do not significantly overlap.

Next, we have addressed the question  whether it is possible to characterize the folding reaction  by averaging over a relatively small number of denatured configurations.
To this end, we have confronted the predictions obtained with the DRP  method with the results obtained by  long MD simulations. We have found that the two methods lead to consistent pictures of the transition. By averaging over the dominant reaction pathways obtained from just 8 different initial conditions, we could accurately predict the location of the valley in the free energy landscape, leading to the native state. 

In the last part of this work, we have investigated the folding mechanism of a $\beta$-hairpin suggested by the simple Go-type model under consideration. We have found that both the MD and DRP results support a hydrophobic collapse mechanism. In fact,  in the initial (final) stage of the unfolding (folding) reaction, the size of the hydrophobic core remains constant and consistent with the value in the native state, while the number of native contacts rapidly drops (increases). 

Based on this work, we  conclude  that the folding reaction of a chain of amino-acids with native interactions and  steric repulsion can be characterized in terms of  just few dominant reaction pathways. This result supports and complements the outcome of a previous validation study of the DRP method, which was performed on a smaller molecule using an atomistic force field \cite{DFP2}.

\acknowledgments
I am grateful to my collaborators H.Orland, F.Pederiva and M.Sega for several important discussions. I thank W.Eaton, G.Hummer and A.Adib, whose questions and comments motivated this work.

\end{document}